\documentclass[fleqn,usenatbib]{mnras}

\usepackage{newtxtext,newtxmath}
\usepackage{graphicx}
\usepackage{amsmath}
\usepackage{ae,aecompl}
\usepackage{graphicx}
\usepackage{xcolor}
\usepackage{ulem}
\usepackage{cancel}
\input epsf.tex

\usepackage[T1]{fontenc}

\DeclareRobustCommand{\VAN}[3]{#2}
\let\VANthebibliography\thebibliography
\def\thebibliography{\DeclareRobustCommand{\VAN}[3]{##3}\VANthebibliography}

\usepackage{graphicx}	

\title[ Irregular Rings ]{Debris Rings from Extrasolar Irregular Satellites} 

\author[Hayakawa and Hansen] {Kevin T. Hayakawa$^{1,2}$\thanks{E-mail: kevin.hayakawa@csuci.edu} and Bradley M. S. Hansen$^{2,3}$\\
$^1$California State University Channel Islands, Camarillo, CA 93012 \\
$^2$University of California Los Angeles, Los Angeles, CA 90095 \\
$^3$Mani L. Bhaumik Institute for Theoretical Physics, Department of Physics \& Astronomy}

\date{Accepted XXX. Received YYY; in original form ZZZ}

\pubyear{2022}

\begin{document}
\label{firstpage}
\pagerange{\pageref{firstpage}--\pageref{lastpage}}
\maketitle

\begin{abstract}
Irregular satellites are the minor bodies found orbiting all four Solar System giant planets, with large semi-major axes, eccentricities, and inclinations. Previous studies have determined that the Solar System's irregular satellites are extremely collisionally evolved populations today, having lost $\sim$99 per cent of their initial mass over the course of hundreds of Myr. Such an evolution implies that the irregular satellites must have produced a population of dusty collisional debris in the past, which is potentially observable due to the resulting reprocessing of stellar light. In this paper we examine the signatures of the debris discs produced by extrasolar analogues of this process. Radiation pressure, quantified by the parameter $\beta$, is the driving force behind the liberation of dust grains from the planetary Hill sphere, and results in the formation of circumstellar dust rings, even in the absence of an underlying belt of asteroids in the system. Our simulated discs reproduce many of the same features seen in some classes of observed debris discs, such as thin ring morphology, a large blowout size, and azimuthal symmetry. We compare our simulated discs' radial profiles  to  those of the narrow dust rings observed around  Fomalhaut and HR 4796A, and show that they can broadly reproduce the observed radial distribution of dust. 
\end{abstract}

\begin{keywords}
stars: abundances -- planets and satellites: formation -- planets and satellites: dynamical evolution and stability -- planets and 
satellites: gaseous planets -- planets and satellites: composition
\end{keywords}




\section{Introduction}



%

Planet formation is a dynamic process, wherein the growth of planets is accomplished via a prolonged history of interactions between smaller bodies, leading to scattering and collision \cite[e.g.,][]{L93}.  This process is particularly important during the latter stages of planetary assembly, as planetary systems settle down into their final configurations. Indeed, the process of dynamical clearing is thought to 
continue for some time after planets have reached their final masses, as the remnants of the source population are ground down and removed from the system \citep[e.g.,][]{GLS04}. Stars in this stage of development often show evidence for extended, tenuous, populations of dust \citep{W08}. These dust grains scatter and re-radiate light from the central star, and can be observed either by looking for infrared excesses or by imaging in scattered light. The lifetime of dust in such systems is short, limited by radiation pressure and Poynting-Robertson drag  \citep[e.g.,][]{BLS79}, but the observation of this material offers essential insights into the architectures of newly formed planetary systems. 
 
As a result, there have been substantial efforts to image such debris systems directly  \citep[see][for a recent summary]{HDM18}. The results show a wide range of morphologies, from extended discs (e.g. those around $\tau$~Ceti and HD~141569) to very narrow rings, such as those around the stars Fomalhaut, HR 4796A, and HD 141011. The variation in appearance presumably indicates some complexity in the evolution and outcome of the planetary assembly process, and there exist detailed models for the kinds of outcomes to expect \citep{W08,K10,KB16,LC16,B21}. 

Debris discs are usually modelled with a source population as a belt of planetesimals undergoing collisional evolution, where the velocity dispersion is stirred either by the development of larger bodies within the belt, or as the result of perturbations from planets in the system \citep[e.g.,][]{W08}. These are natural analogues of the Solar system dust generated either by collisions in the Asteroid belt or the Kuiper belt, although the extrasolar systems are much more massive.

However, there is a third Solar System small body population that is thought to have undergone substantial collisional evolution but has not yet been widely considered in the extrasolar context -- namely the irregular satellites of the giant planets \citep{JH07}. Evolutionary models of this population suggest that it could have been much larger in the past and could have generated a substantial population of dust during the course of losing $\sim 99$ per cent of its original mass \citep{Bott10}. Indeed, such populations are thought to be an inevitable consequence of giant planet formation \citep{NVM07} and the possible existence of irregular satellite clouds around extra-solar planets has recently been postulated to explain the curious properties of the exoplanet candidate Fomalhaut~b \citep{KW10,KB16}.  These papers have focussed on the production of dust near the planet, but radiation pressure forces will cause much of the dust to spiral outwards into a circumstellar ring, and can therefore  also contribute to the observed extended structures observed around many stars.

Therefore, our goal in this paper is to examine the kinds of debris disc signatures one might expect from a source population initially confined to a planetary Hill sphere, and to explore their similarities and differences with those that result from more traditional population assumptions. An alternative to the traditional planetesimal disc model is particularly attractive in the case of the thinnest debris rings, such as those around Fomalhaut and HR 4796A, where the narrowness requires additional  hypotheses such as shepherd satellites \citep{BPC12}, instabilities involving a gaseous component \citep{LK13} or recent collisions \citep{OMT19}. We will demonstrate that irregular satellite clouds naturally give rise to narrow rings and may therefore offer a more natural explanation for these structures, in the sense that the confinement of the original planetesimal population is due to the gravitational influence of the planet.

The outline of this paper is as follows. In \S~\ref{R3B} we describe the dynamics of radiation pressure-driven dust in the reduced three-body problem, and examine the debris disc geometry that results if the source population of the dust is initially restricted to a planetary Hill sphere. In \S~\ref{Source} we then introduce a model for a source population of dust which we combine with the dynamical model to build a model of a candidate debris disc so that we may explore the observational implications of this hypothesis. We then compare these features  to the present state of the art  observations of the two most prominent thin ring systems -- Fomalhaut and HR 4796A -- in \S~\ref{Discuss}.




\section{Dynamics of dust generated in an irregular satellite swarm}
\label{R3B}

The scattering and absorption/re-emission of light in a debris disc is the action of dust particles in orbit about the star. In a traditional debris disc model, this dust is released in collisions between larger bodies in heliocentric orbits, and so reflects the heliocentric distribution of the parent bodies. Here we wish to examine the consequences when the dust is released by collisions between bodies that are localised in orbit around a planet. In addition to the radiation pressure force from the central star, their dynamics is also regulated by the gravitational influence of the planet.

\subsection{Single dust grain dynamics in the restricted three body problem with radiation pressure}

\begin{figure}
\centering
\includegraphics[width=0.5\textwidth]{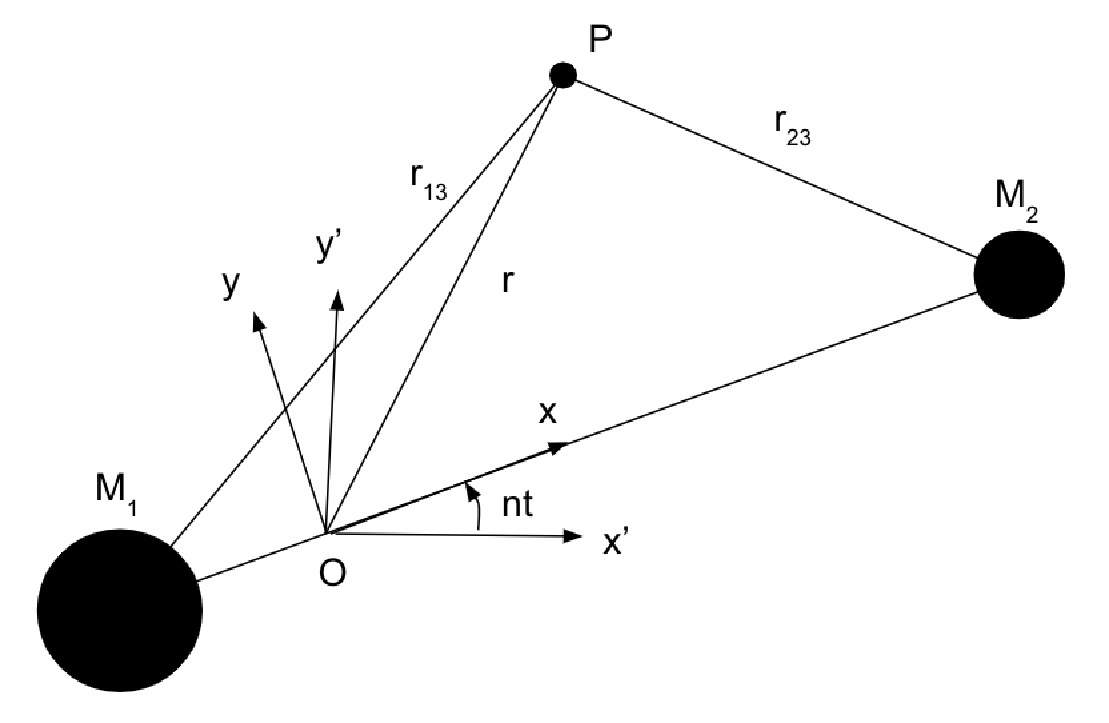}
\caption{\label{3Body} Schematic of the restricted three-body problem. The x- and y-axes make up the corotating centre-of-mass frame, rotating at an angular velocity $n$, centreed at point $O$. The x'- and y'-axes make up the inertial centre-of-mass frame. $M_1$ and $M_2$ are massive bodies with the third body located at point $P$. 
}
\end{figure}


Dust particles have infinitesimal mass, so their dynamics can be treated accurately within the paradigm of the restricted three-body problem, sketched out in Fig. \ref{3Body} \citep{MD99}. However, the stream of photons
emanating from the central star is absorbed or scattered by the dust grains, and exerts a radiation
 pressure. This means that small particles experience a non-negligible additional radial force, which reduces the effective gravity of the central object \citep{Sch80} and fundamentally alters the geometry of the pseudo-potential that regulates the dynamics.
 We can relate this purely radial radiation pressure force to the stellar gravitational force using the formalism of \cite{BLS79} as follows,

\begin{equation}
\textit{\textbf{F}} = -\dfrac{G (1-\beta) M_1 m}{r_{13}^2} \hat{\textit{\textbf{r}}}_{13} -\dfrac{G M_2 m}{r_{23}^2} \hat{\textit{\textbf{r}}}_{23},
\end{equation}

\noindent where $G$ is the Newtonian gravitational constant, $\beta = |\textit{\textbf{F}}_{\textrm{rad}}|/|\textit{\textbf{F}}_{\textrm{grav}}|$ is the relative strength of radiation pressure compared to stellar gravity, $M_*$ is the stellar mass, $m$ is the dust grain mass, $r_{13}$ is the distance from the grain to the star, $r_{23}$ is the distance from the grain to the planet, $\hat{\textit{\textbf{r}}}_{13}$ is the radial unit vector away from the star, and $\hat{\textit{\textbf{r}}}_{23}$ is the radial unit vector away from the planet. For $\beta > 0$, the dust grains behave as if they `see' a star of reduced mass $(1-\beta) M_*$. 

The parameter $\beta$ reflects the strength of the radiation pressure, and can be more precisely quantified as

\begin{equation}
\beta = \dfrac{3L_* \langle Q_{\textrm{rad}} \rangle}{8\pi GM_* c \rho D},
\end{equation}

\noindent where $L_*$ is the stellar luminosity, $\langle Q_{\textrm{rad}} \rangle$ is the wavelength-averaged radiation pressure coefficient, $c$ is the speed of light, $\rho$ is the mass density of grains, and $D$ is the diameter of the grain.
This $\beta$ can be thought of as a proxy for grain size $D$ if we assume a constant mass density $\rho$ among dust grains, since $\langle Q_{\textrm{rad}} \rangle$ is of order unity. \cite{KG01} performed laboratory experiments involving collisions between silicates and found the mass density of resulting grains to be $\rho = 2.8\ \mathrm{g\ cm^{-3}}$. We assume a value for $\langle Q_{\textrm{rad}} \rangle \sim 0.5$ as a rough average from \citet[][]{LS18}. $\beta$ can thus be evaluated as

\begin{equation} \label{beta}
\beta \approx 0.206 \left( \dfrac{D}{1\ \mathrm{\mu m}} \right)^{-1} \left( \dfrac{L_*}{\mathrm{L_\odot}} \right) \left( \dfrac{M_*}{\mathrm{M_\odot}} \right)^{-1},
\end{equation}

\noindent where we have assumed typical values of luminosity and mass of a G-type star such as the Sun.


\begin{figure}
\centering
\includegraphics[width=0.52\textwidth]{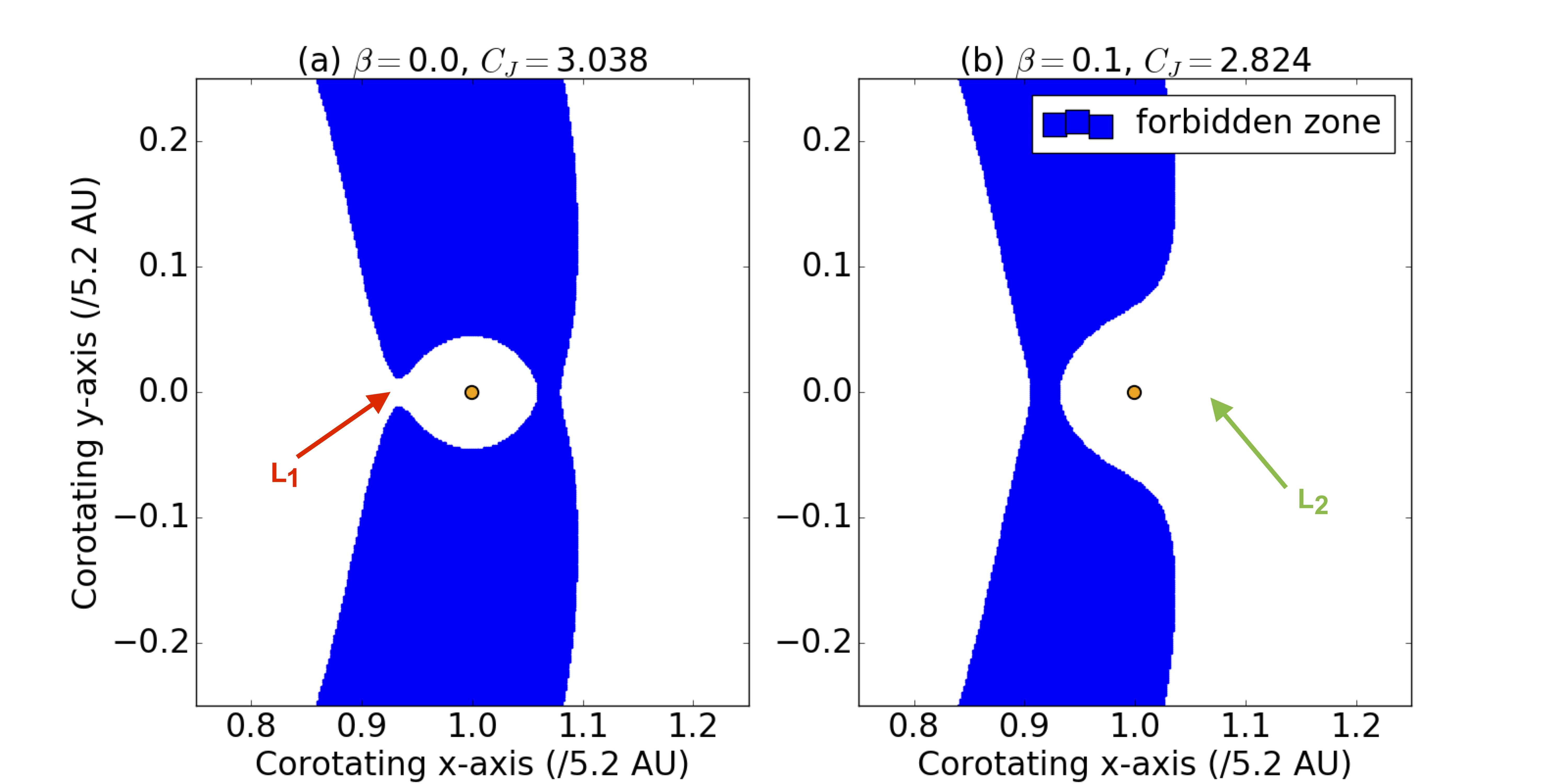}
\caption{\label{Forbidden0} \textit{Panel (a):} forbidden zone (in blue) for a Jacobi constant of $C_J = 3.038$ when radiation pressure is not included $(\beta = 0)$. \textit{Panel (b):} forbidden zone (in blue) for a Jacobi constant of $C_J = 3.824$ when radiation pressure is taken into account $(\beta = 0.1)$. The orange circle (not to scale) represents the location of the giant planet. In panel (a), we note that the Hill spheres of the star and planet overlap at the L$_1$ Lagrange point. Thus, dust grains originating in the planetary Hill sphere are permitted to escape into a circumstellar orbit. In contrast to panel (a), we note that in panel (b), there is now an opening at the L$_2$ Lagrange point for the dust grains to escape from.}
\end{figure}

The dynamics of the test particle in the co-rotating frame of the reduced three-body problem is governed by a pseudo-potential that accounts for both the gravity of the two massive bodies and the centrifugal force \citep{MD99}. The pseudo-potential defines a set of `zero-velocity curves' which restrict the motion of a test particle, depending on its initial conditions.

The fact that the radiation pressure only scales the effective mass of the central star means that the same formalism applies here, but the revision of the coefficients in the pseudo-potential results in an important qualitative difference. Although the direct gravitational force felt by the dust is reduced by the radiation pressure, the orbital velocity of the planet is not similarly affected, and so the relative contributions of the three different components of the pseudo-potential change with $\beta$. In particular, at fixed mass, there is a critical $\beta$ above which the L$_2$ point becomes the global potential minimum (instead of L$_1$, as in the $\beta=0$ case). This distinction is important because it is this minimum that decides the direction in which dust, grinding down in a collisional cascade, leaves the Hill sphere and enters heliocentric orbit. 

To illustrate the effects of this change in geometry, let us  consider two different physical scenarios: one where radiation pressure is not important (i.e., turned `off,' $\beta = 0$) as in panel (a) of Fig. \ref{Forbidden0} and another where radiation pressure is important (i.e., turned `on') as in panel (b) of Fig. \ref{Forbidden0}. Without loss of generality, we take $\beta = 0.1$ for the radiation pressure scenario. If we wish to derive the minimum velocity of escape, we see that, in
 panel (a), particles will more readily escape through the L$_1$ Lagrange point than L$_2$. This behavior is well studied, such as in the case of Roche lobe overflow where mass transfer can occur between two bodies in a binary system. However, when radiation pressure is non-negligible, we see in panel (b) that the lowest velocity particles to escape now overflow L$_2$.
 Thus, the addition of radiation pressure into our equations of motion changes the physics from accretion on to the star to ejection of material outside the orbit of the planet. This is a consequence of the weakened effective gravity of the central star, which shifts its contribution to the pseudo-potential inwards and changes the relative heights of the L$_1$ and L$_2$ points. In Appendix~\ref{Threshold} we review more thoroughly how  this change in topology occurs as $\beta$ is changed.

There are essentially three fates for a dust grain: (i) accretion on to the planet, (ii) accretion on to the star, or (iii) escaping to infinity. Depending on the initial conditions, the dust grain may simply spiral into the planet after an irregular satellite collision, coating the planet, and any satellites, with dark dust and affecting its albedo \citep{BLS79,Bott10}.  Accretion on to the star may occur as the result of  another radiation-related process: Poynting-Robertson (PR) drag. This is a consequence of the loss of angular momentum due to reradiation of absorbed energy by the dust, but is not taken into account here because it occurs on a longer time-scale than the direct dynamical effects of radiation pressure, and generally affects large particles more. We will ignore both PR drag and circumstellar collisions between dust grains in our simulations because their respective time-scales are longer than the ejection time-scales of individual dust grains from the system, as shown in Fig. \ref{PR_drag}. Detailed calculations are performed in Appendix \ref{circumstellar}.

\begin{figure*}
\centering
\includegraphics[width=\textwidth]{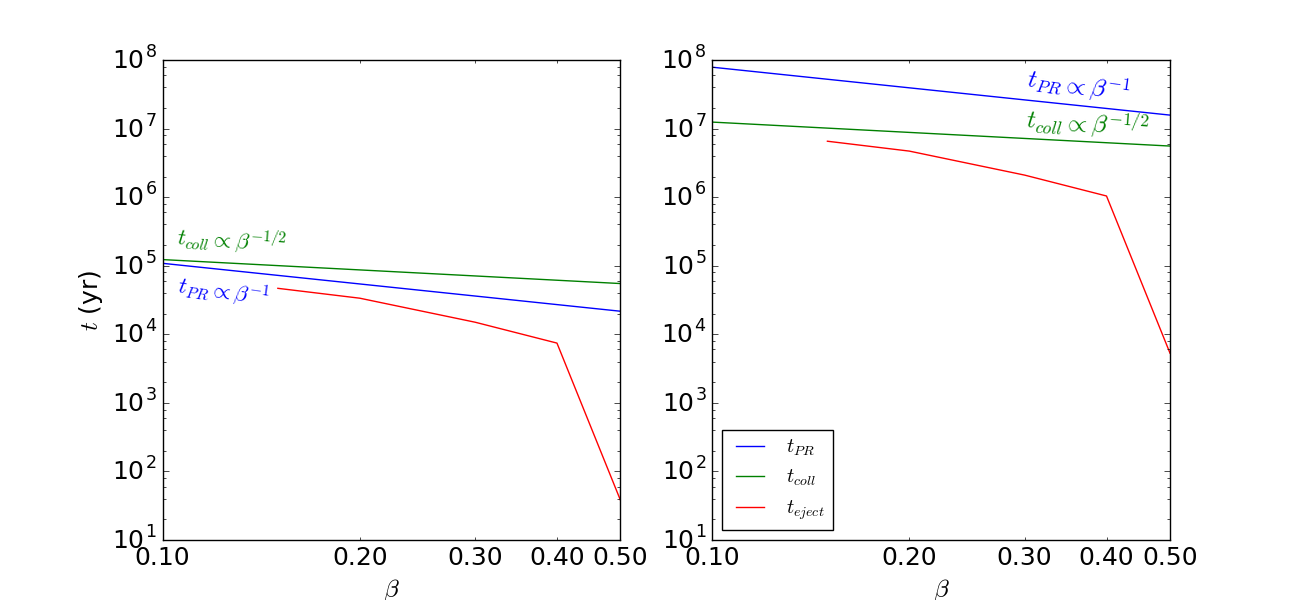}
\caption{\label{PR_drag} PR drag, collisional, and ejection time-scales of circumstellar dust grains as a function of $\beta$, for two canonical examples discussed in this paper. \textit{Left:} For a Jupiter-mass planet orbiting at 5.2 au with $10^{-3}$ $M_L$ of irregular satellites. \textit{Right:} For a Jupiter-mass planet orbiting at 140 au with 1 $M_L$ of irregular satellites. In this paper, we ignore PR drag and circumstellar grain collisions due to their larger time-scales. Collisions cannot always be ignored. We only ignore collisions in this paper because in the situations studied here, the ejection time-scale dominates. Detailed calculations are performed in Appendix \ref{circumstellar}.}
\end{figure*}

For our purposes, the most important outcome is escape from the Hill sphere through the L$_2$ point. Although the eventual outcome is escape to infinity, orbital integrations show that many trajectories spend multiple orbital periods in the vicinity of the outer edge of the relevant zero-velocity curve, before eventually spiralling outwards. This extended period of quasi-regular orbital behaviour thus gives rise to the observational appearance of a thin ring, allied with an exponential tail of orbits that extend out much farther. Such a configuration bears a qualitative resemblance to the `birth ring + halo' model of many debris systems and we will examine its observational consequences below.



\subsection{Sample orbital integrations}\label{sample_orbital_integrations}

 To better understand how this change in geometry reflects itself in orbital behaviour, let us examine the behaviour of a few representative test particles before building a large ensemble population. We perform our numerical integrations using the \textsc{Mercury} N-body integrator \citep[][]{C99}. We start with the simplest case of $\beta = 0$, which represents a particle that is too large for radiation pressure to have an appreciable effect.
Each particle originates in the Hill sphere of its parent planet, and receives a 3-D velocity vector of magnitude $v_\mathrm{i}$, whose direction is oriented randomly.  (We also investigated the effects of preferring prograde or retrograde orbits for our initial conditions, but found no significant change in the results compared to randomly oriented orbits.) After a few orbits around the planet many grains slip through the L$_1$ Lagrange point and `bounce' along the inner edge zero-velocity curve. After several excursions around the star, the grain returns to the Hill sphere through the L$_1$ orbit in a messy, rapidly precessing orbit. In the absence of a dissipative mechanism, this behavior basically repeats itself over time, with the grain being gravitationally shared by the star and the planet. On longer time-scales, Poynting-Robertson drag will eventually decouple the particle from the Hill sphere and it will spiral into the star.

\begin{figure}
\centering
\includegraphics[width=0.48\textwidth]{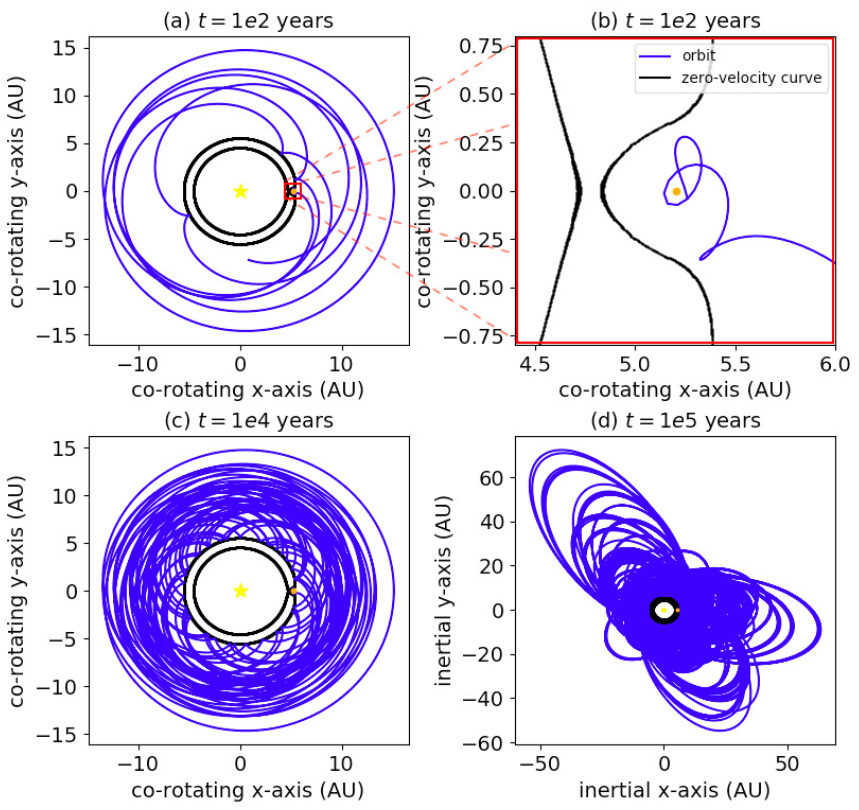}
\caption{\label{B_01} Orbital trajectory of a single 1 $\mu$m dust grain (in blue) overplotted on its respective $\beta = 0.1$ zero-velocity curves (in black). The orbits are shown at various evolutionary stages: $t = 1e2$ yr in panels (a)--(b), $t = 1e4$ yr in panel (c), and $t = 1e5$ yr in panel (d). Panel (b) represents a zoom-in of panel (a).}
\end{figure}

\begin{figure*}
\centering
\includegraphics[width=1.0\textwidth]{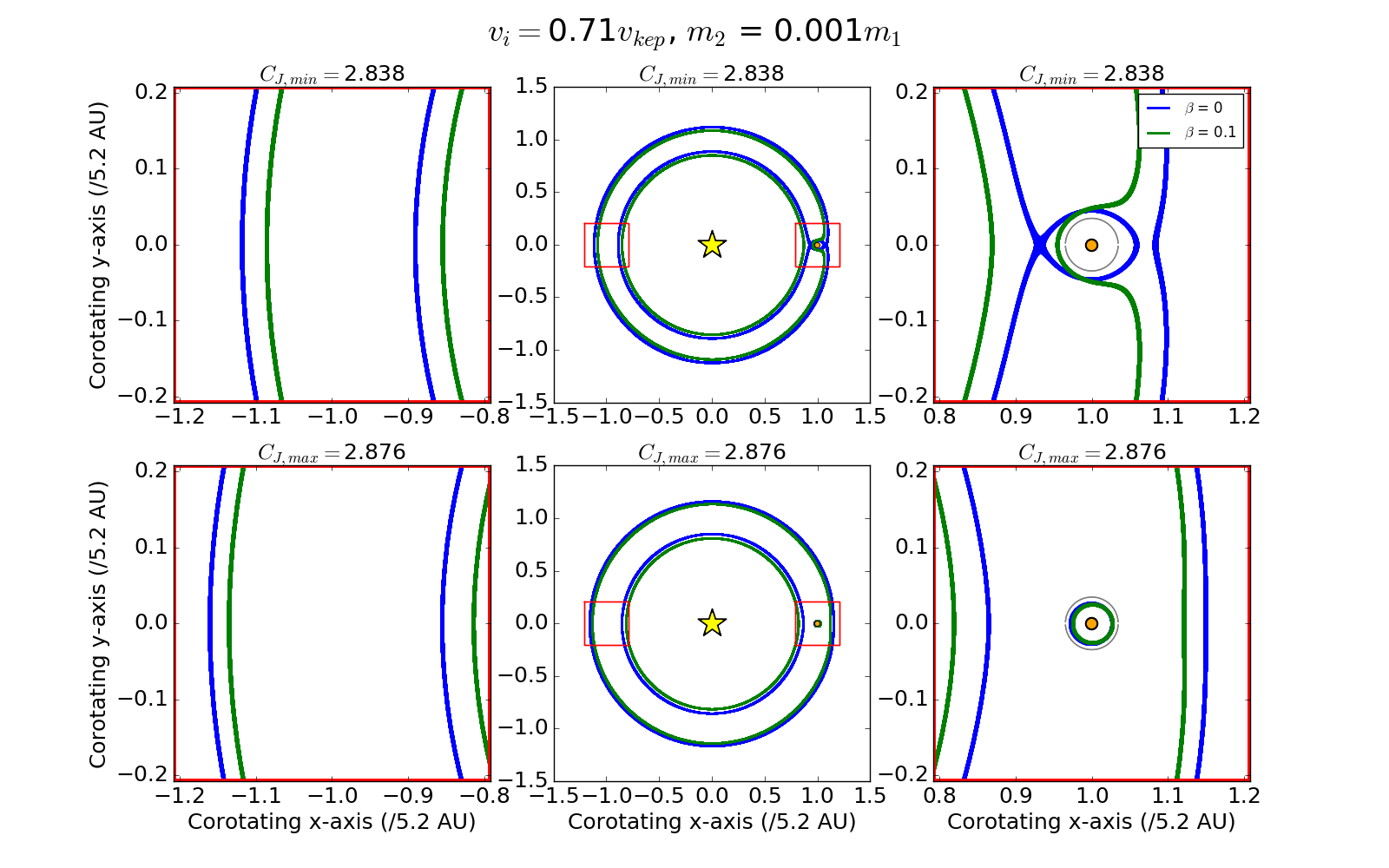}
\caption{\label{forbidden_zone_thickness_example} Zero-velocity curves for both the classic restricted three-body problem ($\beta = 0$) and with moderate radiation pressure included ($\beta = 0.1$), as representative examples for a mixed mass ratio of $M_2/M_1 = 0.001$ and initial velocity $v_\textrm{i}$ equal to 71 per cent of the local circumplanetary Keplerian velocity $v_{\textrm{kep}}$. The left and right columns represent zoomed in versions of the respective red squares shown in the middle column. The top row shows the contours for the minimum Jacobi constant that arise from the initial conditions while the bottom row shows the contours for the maximum Jacobi constant. Thus, all possible Jacobi constants fall between these two extrema. Since the Jacobi constant only depends on the square of the velocity, all possible directions of velocity have implicitly been considered as well as all isotropic orbital configurations. The main two findings here are that the forbidden zone thickness increases with increasing radiation pressure, but the radii of both the inner and outer edges of the forbidden zone shrink with increasing radiation pressure, as seen in the left column. In other words, the radius of the outer edge of the forbidden zone shrinks less than that of the inner edge.}
\end{figure*}


 Next we examine the case of radiation pressure turned `on' with an intermediate strength of $\beta = 0.1$. The path of such a particle is shown in Fig. \ref{B_01}, for the case of a Jupiter mass planet on a circular orbit of semi-major axis 5~au.
 In this case, the particle spirals outwards -- rather than inwards -- and makes
 several cardiod-shaped excursions, roughly several planetary semi-major axes in size,  as shown in panel (a). This is a consequence
 of the aforementioned change in the geometry of the pseudopotential, as shown in panel (b).
  Like in the $\beta = 0$ case, the grain will occasionally come to a sharp halt along the predicted zero-velocity curve. However, the fundamental alteration of  the forbidden zone, caused by the addition of radiation pressure, means that the `collision' occurs with the outer edge of the zero-velocity curve, not the inner one. In panel (c), we see that after a moderate number of dynamical time-scales, this behavior essentially repeats itself since the orbits all stay within $\sim$15 au. However, in panel (d), after a large number of dynamical time-scales, we see that the eccentricity of the grain has been pumped up dramatically, reaching an apoapsis up to $\sim$75 au, until it is effectively ejected from the system.


These sample integrations illustrate that the dynamics of particles released from the planetary Hill sphere under the influence
of radiation pressure can reproduce the basic birth ring configuration of debris disc models, even without the underlying birth ring of planetesimals. We wish now to expand this into a proper model for debris discs. This means we need a more detailed source model, which will link the properties of the dust to the new underlying planetesimal population -- the irregular satellite population. This is the focus of \S~\ref{Source}.

\subsection{Forbidden zone thickness as a function of radiation pressure}

An interesting question is to ask how the results of our simulations will change depending on planet mass, since the study of exoplanets around nearby stars has discovered a great variety in planetary properties. The solutions to the restricted three-body problem depend not just on the mass of the secondary body (the planet), but specifically on the mass ratio of the secondary body to the primary body ($\mu = M_1 / M_2$). Thus, a Saturn-like planet orbiting an M dwarf may have similar dynamics to those of a Jupiter-like planet orbiting a Sun-like star, if both systems have a mass ratio of $\mu \sim 0.001$.  

Increasing radiation pressure generally has the effect of increasing the forbidden zone thickness as seen in the first column of Fig. \ref{forbidden_zone_thickness_example}. However, while the overall thickness increases, we can see that the radii of both the inner and outer edge of the forbidden actually decrease. In other words, the radius of the outer edge is decreasing by a smaller amount than that of the inner edge.

For a more comprehensive look at how the forbidden zone changes as a function of both radiation pressure strength and mass ratio, interested parties may refer to the discussion in Appendix \ref{forbidden_zone_thickness_appendix}.

\section{Generation of dust in an irregular satellite swarm}
\label{Source}

We assume that the particles  whose orbits we track originate from collisions between irregular satellites orbiting around the giant planet. Irregular satellites revolve around their parent planet at relatively large distances compared to other moons \citep[e.g.,][]{Bott10}, so it is natural to characterize their distances in units of Hill radii, given by $R_\mathrm{H} = a_\mathrm{p}[M_\mathrm{p}/(3M_*)]^{1/3}$, where $a_\mathrm{p}$ is the planetary semi-major axis and $M_\mathrm{p}$ is the planetary mass. The original orbits of irregular satellites are believed to be roughly isotropically distributed \citep{JH07}, so we investigate both prograde and retrograde orbits around the parent planet, which are typically found at $r_{23}/R_\mathrm{H}$ values of $\sim 0.1$ to $0.5$, where $r_{23}$ is the distance between the planet and the dust grain. Thus, we use those upper and lower limits to randomly generate starting locations for the dust grains.

We divide the discussion of initial velocities into magnitude and direction. We take the magnitude of the velocity to be 71 per cent of the Keplerian circular velocity at the debris's respective distance from the planet. Since it is a spherically symmetric cloud, we assume that the direction of the dust grain's velocity unit vector is random. Specifically, in polar coordinates $\theta$ and $\phi$, $\cos(\theta)$ is distributed uniformly in [-1,1) and $\phi$ is distributed uniformly in [0,2$\pi$).  We find no significant difference  between  the qualitative results  for orbits that are initially prograde or retrograde. Since the Keplerian velocity is given by $v_{\textrm{kep}}=(G M_\mathrm{p} / r_{23})^{1/2}$, the initial velocities are given by

\begin{equation}
    v_{\textrm{kep}} = (2.248\ \mathrm{km\ s^{-1}}) \left( \dfrac{M_\mathrm{p}}{\mathrm{M_J}} \right)^{1/2} \left( \dfrac{r_{23}}{0.5\ R_\mathrm{H}} \right)^{-1/2}
\end{equation}

\noindent where $\mathrm{M_J}$ is the mass of Jupiter.


\subsection{Rate of dust generation}

A population of irregular satellites will generate a collisional cascade, 
in which planetesimals are ground down to micron-sized dust grains. Collisions between the largest collisionally coupled bodies of size $D_{\mathrm{max}}$ initiate the cascade, creating numerous medium-sized bodies that further collide with each other to produce successively smaller bodies.
In the traditional context of a circumstellar debris disc, the smallest collisionally coupled body's size $D_{\mathrm{min}}$ is determined by the strength of the central star's radiation pressure, and tends to be about 1 micron. This is often referred to as the blowout size. \citet[][]{D69} found that a self-similar, steady-state cascade follows a power law differential size distribution governed by

\begin{equation}\label{Dohnanyi}
    \dfrac{dN}{dD} \propto D^{-q},
\end{equation}

\noindent where $D$ is the spherically symmetric grain's diameter and $q \approx 3.5$.
 A dust grain is no longer in a bound orbit around the star when the ratio of radiation pressure force to gravitational force is greater than 0.5 \citep[e.g.,][]{PK15}, i.e.,

\begin{equation}
    \beta \equiv \dfrac{F_{\mathrm{rad}}}{F_{\mathrm{grav}}} \ge 0.5.
\end{equation}

\begin{figure}
\centering
\includegraphics[width=0.48\textwidth]{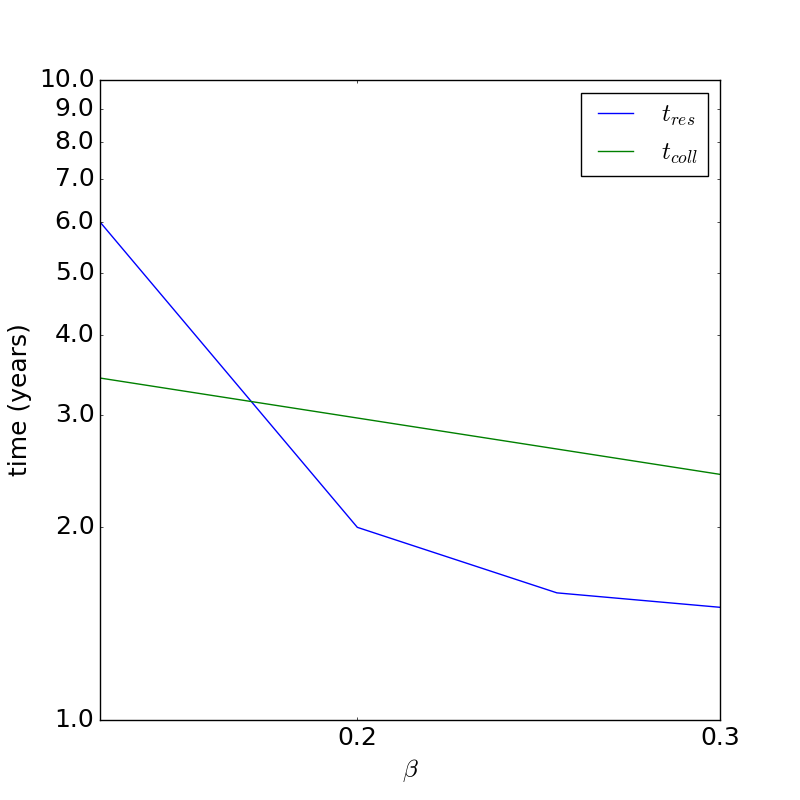}
\caption{\label{t_res} Characteristic dust grain residence time in the Hill sphere (in blue) -- derived from numerical integration of orbits -- and catastrophic collisional time-scale (in green) -- from equation~(\ref{Colltime}), both as functions of $\beta$ {when $M_\text{tot} = 1\ M_\text{L}$ and $\rho = 3$ g/cm$^3$, with the other parameters being described by Eq. \ref{Colltime}}. The intersections of the curves at approximately $\beta = 0.18$ represents the critical size at which a dust grain can decouple from the collisional cascade and be expelled from the Hill sphere.}
\end{figure}

{\bf }In the case discussed here, there is an additional consideration. Fragments from irregular satellite collisions will continue to participate in the collisional cascade as long as they orbit within the planetary Hill sphere. However, once the radiation pressure is strong enough for the particle to escape the Hill sphere, the density of collision targets drops dramatically and the collisional time-scale becomes large. Thus, the minimum particle size in the cascade is set by the  size for which the residence time within the Hill sphere  equals the characteristic collision time for particles of that size.
 The residence time here is defined as the amount of time a dust grain spends in the Hill sphere at a given $\beta$ before escaping.
 Conversely, this also sets a minimum $\beta$ for the particles in the more extended debris disc and thus regulates their size.

We can find the collisional time-scale $t_{\textrm{coll}}$ for any member of the collisional cascade from $t_{\textrm{coll}} = 1/(n \sigma v_{\textrm{rel}})$, where $n$ is the number density of particles that cause catastrophic collisions, $\sigma$ is the cross section, and $v_{\textrm{rel}}$ is the relative velocity between impactors. The number density of particles is given by $n = N / V$, where $N$ is the number of particles and $V$ is the volume they occupy. We calculate $N$ by integrating Eq. \ref{Dohnanyi} from $D/2$ to $2D$, the range by which we define sizes that are capable of a catastrophic collision. Additionally, we normalize Eq. \ref{Dohnanyi} by integrating the collisional cascade over mass via

\begin{equation}
    M_{\mathrm{tot}} = \int_{D_\mathrm{min}}^{D_{\mathrm{max}}} m \dfrac{dN}{dD} dD
\end{equation}

\noindent where $m = (4\pi/3)(D/2)^3 \rho$ is the mass of a body in the cascade. Since the irregular satellites are distributed isotropically in a spherical cloud, we take this volume to be some spherical shell with the radius the fraction $f_{\mathrm{tot}}$ of the Hill sphere they occupy: $V = f_{\mathrm{tot}} V_H = f_{\mathrm{tot}} (4\pi/3)R_\mathrm{H}^3$. Gravitational focusing is not important for submillimeter-sized particles, the cross section is just the geometric cross section: $\sigma = \pi(D/2)^2$. Lastly, we take the relative velocity $v_{\textrm{rel}}$ to be of order the circumplanetary Keplerian velocity $v_{\textrm{kep}}$ since the orbital inclinations of irregular satellites are randomly oriented. Putting everything together, we find that the collisional time-scale is

\begin{equation}
    \begin{aligned}
        t_{\textrm{coll}} =\ & (726\ \mathrm{yr}) \left( \dfrac{\beta}{0.1} \right)^{-1/2} \left( \dfrac{M_{\mathrm{tot}}}{10^{-2} \mathrm{M_L}} \right)^{-1} \left( \dfrac{\rho}{1\ \mathrm{g\ cm^{-3}}} \right)^{1/2} \\
        & \times \left( \dfrac{D_{\mathrm{max}}}{30\ \mathrm{km}} \right)^{1/2} \left( \dfrac{M_*}{\mathrm{M_\odot}} \right)^{-5/3} \left( \dfrac{M_\mathrm{p}}{\mathrm{M_J}} \right)^{-2/3} \left( \dfrac{L_*}{\mathrm{L_\odot}} \right)^{1/2} \\
        & \times \left( \dfrac{\langle Q_{\mathrm{rad}} \rangle}{0.5} \right)^{1/2} \left( \dfrac{a}{{\mathrm{a_J}}} \right)^{7/2} \left( \dfrac{f}{0.4} \right)^{1/2} \left( \dfrac{f_{\mathrm{tot}}}{0.098} \right),   \label{Colltime}
    \end{aligned}
\end{equation}

\noindent where $M_{\mathrm{tot}}$ is the total mass of the collisional cascade, $\mathrm{M_L}$ is lunar masses, ${\mathrm{a_J}}$ is the semi-major axis of Jupiter, $f$ is the orbital radius of the body as a fraction of the Hill radius ($f = r_{23} / R_\mathrm{H}$).
As long as the residence time $t_{\textrm{res}}$ is larger than the collisional time-scale $t_{\textrm{coll}}$, we expect the grains to continue to grind down to smaller sizes, which
increases $\beta$ and shortens the residence time-scale. We empirically measure the residence time from our simulations, specifically defining $t_{\textrm{res}}$ at a given $\beta$ to be the amount of time it takes for 50 per cent of the dust grains to exit the Hill sphere. A particle is defined as having exited the Hill sphere if the planetocentric distance $r_{23} > R_\mathrm{H}$.


Fig.~\ref{t_res} shows the comparison of  characteristic collisional and residence time-scales for the dust generated in irregular satellite collisions for a Jupiter-like planet orbiting a Sun-like star {when $M_\text{tot} = 1\ M_\text{L}$ and $\rho = 3$ g/cm$^3$, with the other parameters being described by Eq. \ref{Colltime}}. At low $\beta$ (large particles), the residence time within the Hill sphere is long, because radiation pressure is weak, but as $\beta$ increases, the residence time falls sharply as the radiation pressure accelerates the exodus. Although the collision time also gets shorter with decreasing size, the dependence is flatter. The net result is that the cascade to smaller size is truncated when $\beta$ is large enough that the particles exit the Hill sphere before undergoing any more collisions. We identify the critical $\beta = 0.18$ from the intersection of the $t_{\textrm{res}}$ and $t_{\textrm{coll}}$ curves. This critical $\beta$ represents the largest size dust grain that could escape from the Hill sphere. In principle, the size of an escaping particle could be lower or higher, depending on the parameters assumed in Equation \ref{Colltime} to calculate the collisional time-scale. While it's possible for the residence time (blue curve in Fig. \ref{t_res}) and collisional time-scale (green line) to never intersect depending on the assumed parameters, particles could still escape from the Hill sphere. A short collisional time-scale just means that particles will grind down all the way to the classical ``blow-out'' size corresponding to approximately $\beta = 0.5$ before getting ejected out of the entire system \citep[][]{K10}. {We note here that the classical blowout size of $\beta = 0.5$ only applies to a dust grain on a purely circumstellar orbit, and should only be used as a general benchmark for systems like ours where there is a planet involved.}


\subsection{Source Model}

Our integrations yield a large number of trajectories as a function of time, for different $\beta$. A realistic model assumes that individual dust grains are generated at a constant rate due to the collisional cascade initiated by the irregular satellite population. In practice, we achieve this continuous generation  by offsetting the initial time of individual grain trajectories from our library of integrations. Working within the co-rotating frame ensures that new dust grains will always be produced in the Hill sphere of the planet. We then track the dynamical evolution of these dust grains to calculate density profiles of the dust population, as described in the next section.



\section{Results}

The irregular satellite source model for generating dust discs is fundamentally different from the traditional planetesimal disc source population in that it has a localised source region, confined within the planetary Hill sphere, from which the material spreads out slowly. In this section we wish to therefore characterise the observational appearance of the resulting dust population, which we term an ISDD (Irregular Satellite Debris Disc).


\subsection{From circumplanetary to circumstellar orbits}

\begin{figure*}
\centering
\includegraphics[width=\textwidth]{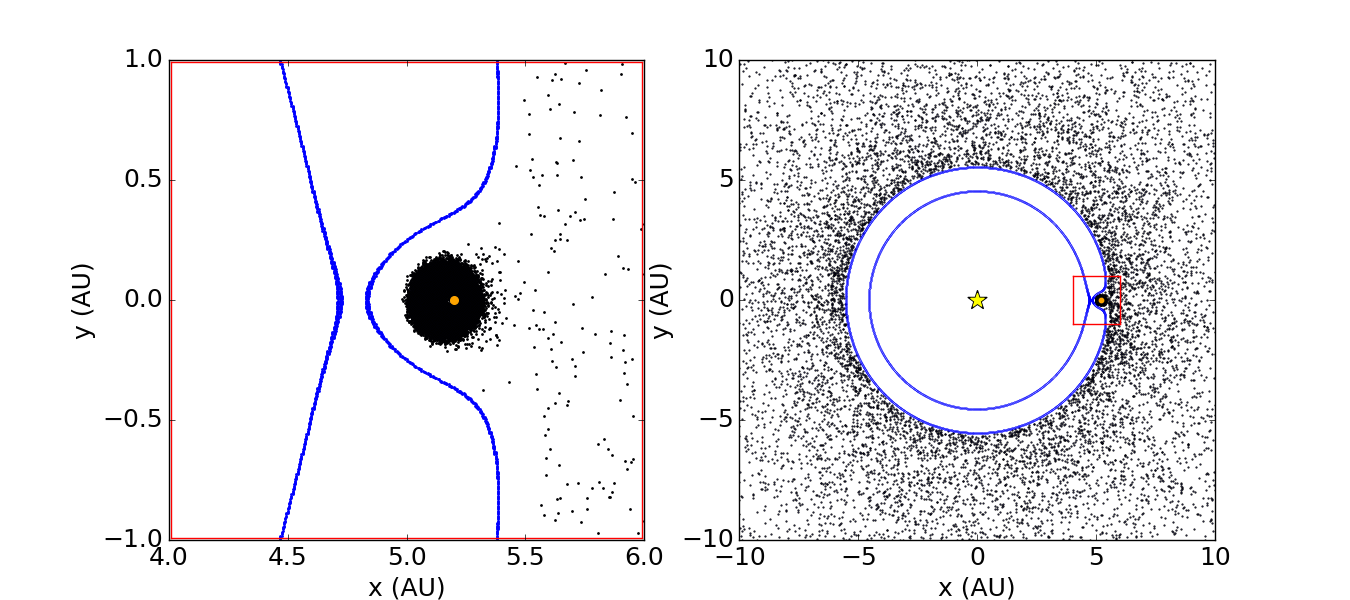}
\caption{\label{synthetic_image}   Position of test particles moving under the influence of radiation pressure for a value of $\beta = 0.15$ after 1,000 dynamical time-scales ($\sim$12,000 years) of integration. This simulation is for a Jupiter-mass planet on a 5.2 au orbit around a solar-mass star. This simulation duration was chosen because it is the time-scale over which a steady state is reached for the shape of the radial profile of the disc. Once a steady state is reached in the disc, the shape of the radial profile remains the same, but the amplitude decreases. The zero-velocity curves for $C_J = 2.824$ are plotted in blue. The location of the star and planet are denoted by a yellow star and orange dot respectively. The right panel shows an face-on view of the disc as a whole. The left panel zooms in on the vicinity of the planet, showing  an overdensity of material in the Hill sphere.  These are particles whose trajectories remain bound over the course of the simulation. In reality, these particles have orbited the planet for hundreds of collisional times and will have been ground down to much smaller sizes. Therefore, it
 is important to note that the overdensity within the Hill sphere is not physical, and is shown here primarily to illustrate the nature of the dust trajectories and their evolution. When viewed as a synthetic image, the panel on the right should have the circumplanetary excess reduced by a factor of several hundred, at least.
  }
\end{figure*}

Fig.~\ref{synthetic_image} shows a snapshot of the positions of $N = 750,000$ particles with $\beta=0.1$, integrated for $10^3$ planetary dynamical times ($\sim$12,000 yr). This simulation is for a Jupiter-mass planet on a 5.2 au orbit around a solar-mass star. This simulation duration was chosen because it is the time-scale over which a steady state is reached for the shape of the radial profile of the disc. The distribution of dust grains can be divided into circumstellar material and circumplanetary material. The circumstellar material is made up of dust grains that were able to escape from the Hill sphere whereas circumplanetary material represents grains that are still trapped in the Hill sphere. Whether or not a dust grain successfully escapes from the Hill sphere is primarily determined by initial conditions. The role of  the zero-velocity curves from the Jacobi formalism in shaping the distribution of the escaped material is clear. Aspects of this dust population, such as azimuthal symmetry, radial profile, ring thickness, and vertical profile will be examined in the following subsections.

It is important to note that the overdensity within the Hill sphere in Fig.~\ref{synthetic_image} is
artificial  because those particles, that do not escape, will be subjected to continued collisional evolution not included in this simulation, and  the particles are shown here primarily to illustrate the nature of the dust trajectories and their
evolution.
 As we previously saw in Fig. \ref{t_res}, the intersection of the residence time and collisional time-scale occurs at a very short time-scale (<100 yr). This is very short compared to the duration of the simulation ($\sim$12,000 yr), so we expect trapped dust grains to still be coupled to the original cascade, grinding down to even smaller sizes until they too are blown out of the Hill sphere. As a result, we would not expect to see a pile-up of material in the form of a circumplanetary disc.



\subsection{ISDDs are azimuthally symmetric}

\begin{figure}
\centering
\includegraphics[width=0.5\textwidth]{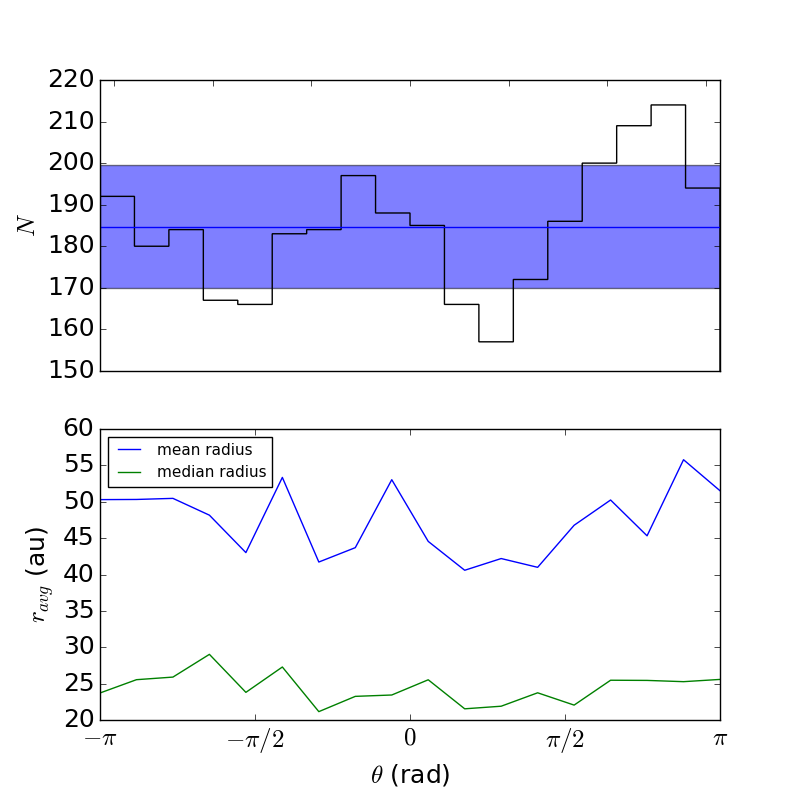}
\caption{\label{az_prof_wide_B_0.1.eps} \textit{Top:} Histogram of dust grain distribution as a function of azimuthal angle $\theta$ for $\beta = 0.15$. The spike located at $\theta = 0^{\circ}$ has been subtracted off, since leftover material in the planetary Hill sphere is not part of the circumstellar disc. The average number of dust grains per bin is 185 with a standard deviation of only 15 grains, denoted by the horizontal blue line and blue shaded region. \textit{Bottom:} Mean and median radius as a function of azimuthal angle $\theta$.  The mean radius is approximately 50 au at nearly all angles while the median radius is approximately 25 au at all angles, showing that the disc is very azimuthally symmetric. }
\end{figure}

Although the dust is initially generated within the Hill sphere, once it escapes, azimuthal symmetry in an ISDD is achieved very quickly. Fig. \ref{az_prof_wide_B_0.1.eps} shows the azimuthal profile of the dust after 100 dynamical time-scales, separated into 20-degree bins for the case of $\beta=0.15$. We have intentionally excluded material that remains in the planetary Hill sphere so that the global circumstellar disc features could be examined.

We quantify the baseline fluctuations in the azimuthal profile by comparing the mean value to the standard deviation. We conclude for four representative values of $\beta$ in the range $0.15 \leq \beta \leq 0.30$ that the ISDDs are azimuthally symmetric since the standard deviations are small compared to the mean. For example, in the case of $\beta = 0.15$, the average number of dust grains per bin is 185 while the standard deviation is 15, so we conclude that the variations are small. Similar results are obtained for other  values of $\beta$. Dust rings generated in this manner nevertheless retain the appearance of azimuthal symmetry.

Another way to evaluate the azimuthal symmetry of the disc is to calculate the average dust grain semi-major axis as a function of azimuthal angle $\theta$. In Fig. \ref{az_prof_wide_B_0.1.eps}, we calculate both the mean and the median radius for the $\beta = 0.15$ disc.  We find that the mean radius (in blue) is approximately 50 au and that the median radius (in green) is approximately 25 au.  It is not surprising to see that the mean radius is higher than the median radius. While the vast majority of dust grains spend their time close to the planet's orbit bouncing around the edges of the Jacobi contours, a small fraction of dust grains will slowly diffuse out of the system due to the influence of radiation pressure and stirring from the planet, biasing the mean to higher radii.

\subsection{ISDDs exhibit thin ring morphology}

Astronomers commonly quantify ring thickness as the ratio of ring width to ring radius $\Delta R/R$. Specifically, a ring may be characterized as `thin' if $\Delta R/R < 0.5$ {\citep[][]{HDM18}}. This definition takes into consideration the great diversity of size scales that debris discs are observed to span and allows us to compare  systems on large and small absolute scales. However, the ratio requires us to specify how we define $\Delta R$ and $R$. We first fit a function to the distribution and find that several of the parameters naturally characterize the ring width and ring radius. 

\begin{figure}
\centering
\includegraphics[width=0.5\textwidth]{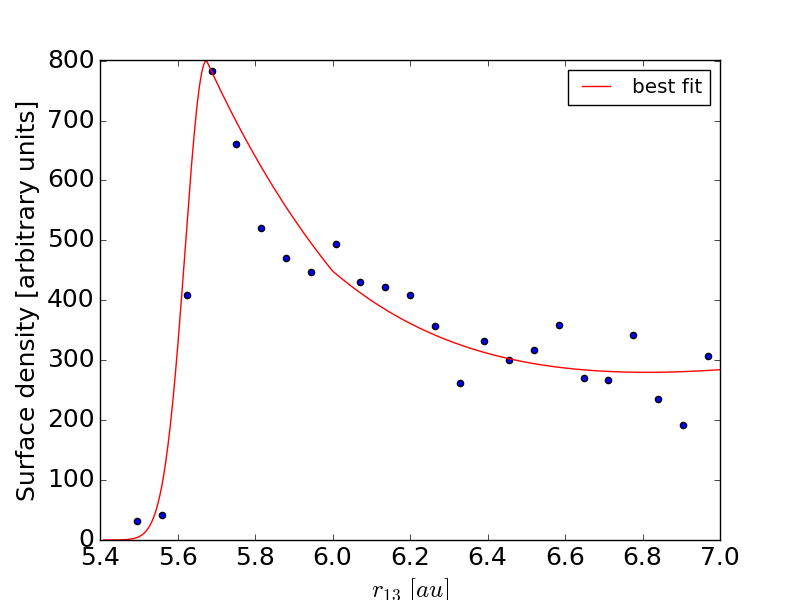}
\caption{\label{rad_prof_B_0_15} Surface density profile for a system with a 1 $\mathrm{M_\odot}$ star and a $10^{-3}\ \mathrm{M_\odot}$ planet  as a function of radius for $\beta = 0.15$. Function fit (in red) to the distribution. The radii that define the boundaries between the three regimes are $r_A = 5.67$ au and $r_B = 5.98$ au. There is a sharp inner edge, indicating the existence of a gap in the disc created by the forbidden zone predicted from the restricted three-body problem. There is a more shallow decline at large distances caused by dust grains taking their time spiraling out of the system due to stellar radiation pressure. The characteristic lengths of the Gaussian and exponential fits allow us to quantify the width of the ring. These width measurements can be compared to observations since only the brightest, densest regions would be observable. For this particular value of $\beta = 0.15$, we find the normalized ring width to be $\dfrac{\Delta R}{R} = 0.148\ _{-0.025}^{+0.023}$, thus meeting the criterion for a thin ring.}
\end{figure}

Specifically, we fit a piecewise function to the radial profile that contains three physically motivated regimes.
Region I ($r_{13} < r_\mathrm{A}$) is simply a one-sided Gaussian that describes the sharp inner edge of the ring. This feature is to be expected since the forbidden zones from the Jacobi contours prevent dust grains from wandering any closer to the star than one planetary semi-major axis plus one Hill radius ($a + R_\mathrm{H}$). Region II ($r_\mathrm{A} < r_{13} <  r_\mathrm{B}$) is an exponential decay function that describes the initial drop off in surface density that occurs as we move outward away from the peak. The peak tends to be just outside of the Jacobi contours because the dust grains spend a lot of time bouncing around the edges of the zero-velocity curves before they diffuse out of the system. Lastly, Region III ($r_{13} > r_\mathrm{B}$) is a continuation of Region II with an added exponential term to soften the drop-off and match the more gradual decay of the outer edge. This feature is also expected since we are investigating moderate radiation pressure strengths ($\beta = 0.15-0.30$) that are strong enough to perturb the dust grains from a circumplanetary orbit to a circumstellar orbit, but not strong enough to immediately eject the grains from the system. The gradual tail of the radial distribution represents dust grains that are in the process of slowly diffusing out of the system. A sample fitted radial profile for $\beta = 0.15$ is shown in Fig. \ref{rad_prof_B_0_15}. The resulting functional form is 

\begin{equation}
    N(r_{13}) = \left\{
        \begin{array}{ll}
            \dfrac{N_0}{r_{13}} \exp \left( -\dfrac{(r_{13} - r_\mathrm{A})^2}{2\sigma_1^2} \right), & r_{13} \leq r_\mathrm{A} \\
            \dfrac{N_0}{r_{13}} \exp \left( -\dfrac{(r_{13} - r_\mathrm{A})}{\sigma_2} \right), & r_\mathrm{A} < r_{13} < r_\mathrm{B} \\
            \dfrac{N_0}{r_{13}} \exp \left( -\dfrac{(r_{13} - r_\mathrm{A})}{\sigma_2} \right) \\
            +\ \dfrac{N_1}{r_{13}} \left[ 1 - \exp \left( -\dfrac{(r_{13} - r_\mathrm{B})}{\sigma_3} \right) \right], & r_{13} > r_\mathrm{B}  
        \end{array}
    \right.
\end{equation}

\noindent where $N_0$ and $N_1$ are normalization constants for their respective terms, $r_\mathrm{A}$ is the peak of the distribution, $r_\mathrm{B}$ is the transition point between Regime II and Regime III, $\sigma_1$ is the standard deviation of the single-sided Gaussian, and $\sigma_2$ and $\sigma_3$ are the characteristic lengths of their respective exponential decay terms.

In order to cast this in the observational variables  $\Delta R$ and $R$, we take $R \equiv r_\mathrm{A}$ since it naturally describes the peak of the distribution, and $\Delta R \equiv \sigma_1 + \sigma_2$ since those lengths each characterize the drop-off in either direction away from the peak. Thus, in terms of the function parameters, $\Delta R / R \equiv (\sigma_1 +\sigma_2)/r_\mathrm{A}$.

\begin{figure}
\centering
\includegraphics[width=0.5\textwidth]{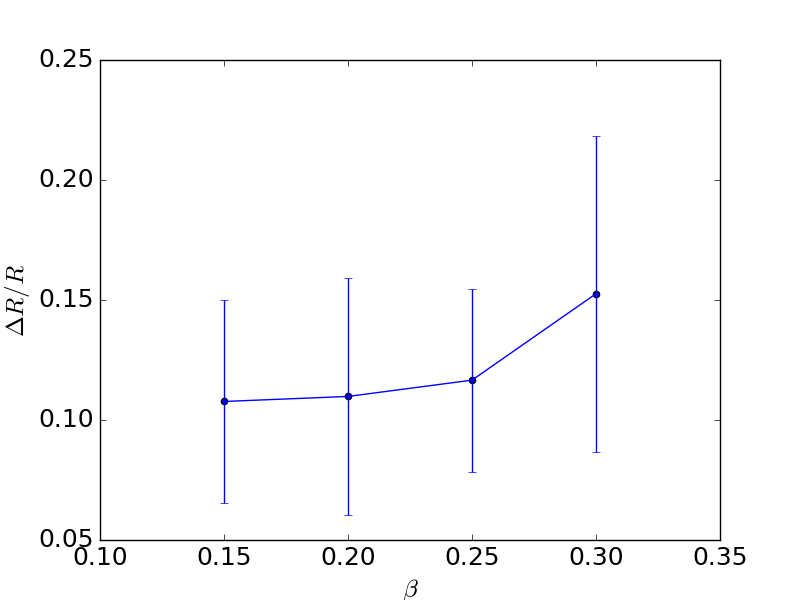}
\caption{\label{FWHM}  Normalized ring widths ($\Delta R / R$) for various values of $\beta$  for a system with a 1 $\mathrm{M_\odot}$ star and a $10^{-3}\ \mathrm{M_\odot}$ planet. The general trend points to broader ring widths and therefore shallower drop-offs for higher values of $\beta$. This is not surprising since higher values of $\beta$ correspond to stronger radiation pressure. Thus, broader ring widths show dust grains that are actively spiraling out of the system. We also note that the error bars tend to be larger for larger values of $\beta$. This is also not surprising since these discs tend to have smaller sample sizes by the end of a controlled simulation, since the stronger radiation pressure ejects a higher percentage of grains from the system.}
\end{figure}

We also apply the fit to multiple homogeneous discs of different values of $\beta$ ($\beta$ = 0.15 -- 0.30). The same piecewise function was fitted to all simulations and for each value of $\beta$, the piecewise function did a good job of smoothly connecting the surface density profile and defining a reasonable ring width. We measure their normalized ring widths and determined the uncertainty by marginalizing out the seven parameters in the function described above in favor of $\Delta R / R$. A $3\sigma$ confidence interval was used to determine the upper and lower limits of the ensuing error bars. Those results are summarized in Fig. \ref{FWHM}. The general trend is that the thickness of the ring grows with increasing $\beta$. This occurs because higher values of $\beta$ correspond to stronger radiation pressure. This stronger pressure is able to more efficiently push dust grains into larger and more eccentric orbits.

\subsection{An exponential tail}

\begin{figure}
\centering
\includegraphics[width=0.5\textwidth]{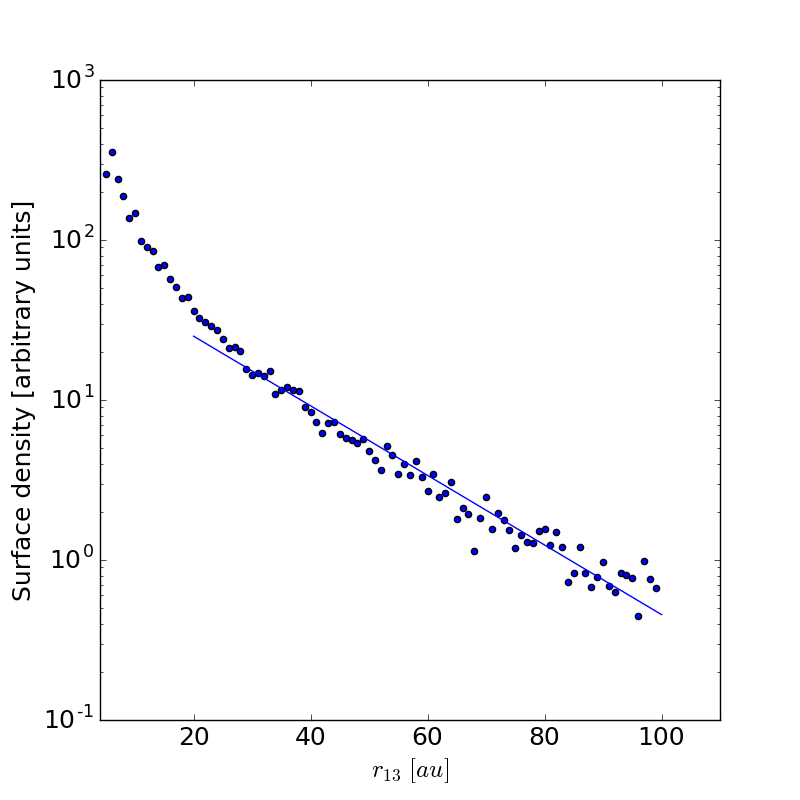}
\caption{\label{exponential_tail} Semilog radial profile of fiducial ISDD for $\beta = 0.15$ showing the entire disc from 0 to 100 au. An exponential decay function (blue line) was fit to the portion of the data from 20 to 100 au to determine the characteristic length of the decay. We find that the profile has a characteristic decay scale of 20.0 au.}
\end{figure}

%
As mentioned in the previous section, the radial distribution has a gently sloped exponential tail. This tail is a generic feature of all models that take radiation pressure into account. However, our model generate larger particles than a standard source model since the collisional cascade is truncated by the Hill sphere residence time, as shown in Fig. \ref{t_res}. The slope of the fiducial $\beta = 0.15$ ISDD is characterized in Fig. \ref{exponential_tail}. While the data plotted is from 0 to 100 au, we calculated the slope only for the portion between 20 and 100 au. The characteristic length came out to be 20.0 au, several times the planet semi-major axis, indicating  a relatively slow drop-off.

\subsection{ISDDs exhibit a toroidal shape}

\begin{figure}
\centering
\includegraphics[width=0.5\textwidth]{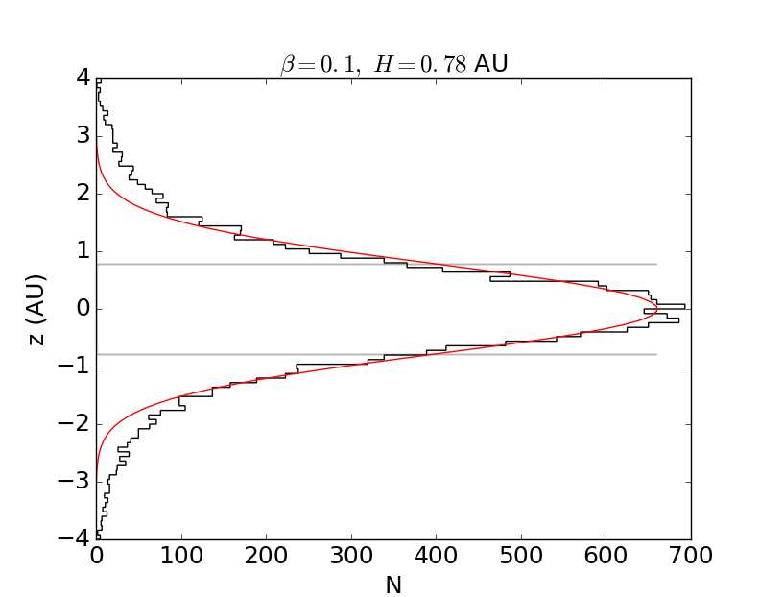}
\caption{\label{vert_dist}  Distribution of dust grains as a function of height $z$. A Gaussian function (in red) has been fit to the distribution. The gray horizontal lines represent one scale height ($H = 0.78$ au) above and below the midplane. }
\end{figure}

We examined the vertical structure of the simulated disc in addition to the radial structure. Specifically, we plotted the dust grain abundance as a function of height $z$ above or below the midplane of the disc, as seen in Fig. \ref{vert_dist}. In addition, we fit a standard Gaussian function to the vertical distribution since the standard deviation would naturally translate to a scale height. We find that the scale height of $H = 0.78$ au for the $\beta = 0.1$ toy model is comparable to both the ring thickness ($\Delta R = 0.64$ au) and Hill radius ($R_\mathrm{H} = 0.35$ au).

We attribute the sharp inner edge of the torus to the forbidden zone predicted by the Jacobi constant. Recall from Fig. \ref{Forbidden0} that particles of a certain Jacobi constant are not allowed to exist in certain spaces in the restricted three-body problem. For physically relevant values of $\beta$, this region takes the shape of an annulus along the orbit of the planet, approximately one Hill diameter in width. The inner edge of our ISDD represents dust grains bouncing around the edges of the forbidden zone.


\section{Comparison to Observations}
\label{Discuss}


 One motivation for this study is the presence of narrow ring-like structures in some observed debris disc systems. The narrowness of the images implies a mechanism for confining either the dust or its parent population. In our model, this is a consequence of the orbital evolution regulated by the Jacobi constant, which sets a hard inner edge on the distribution. The outer profile is more gradual as the dust spirals out, and we compare here this theoretical expectation to properties of the best quantified observed systems. 

\subsection{Fomalhaut}

\begin{figure}
\centering
\includegraphics[width=0.45\textwidth]{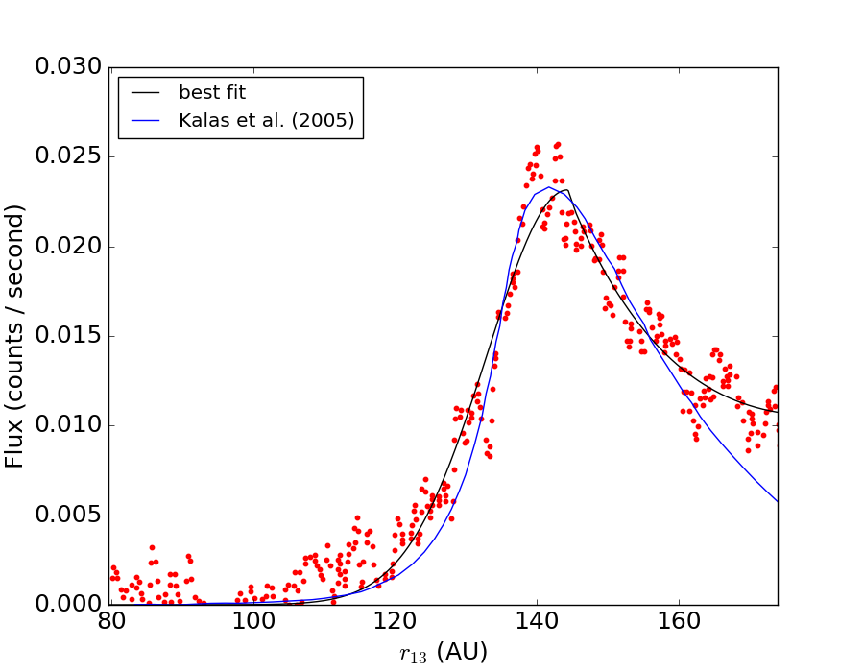}
\caption{\label{kalas}  Fomalhaut flux profile as a function of radius \citep[][]{K05}. The red points are raw observational data. The fit of \citet[][]{K05} is in blue, and our functional fit is in black. They showed that the inner edge of the belt can be modeled as either a power law fit with an index of $\alpha = 10.9$ or an exponential growth proportional to $\exp(0.08r)$, where $r$ is in units of au. Additionally, they showed that the outer edge of the belt can be modeled as either a power law fit with an index of $\alpha = -4.6$ or an exponential decay proportional to $\exp(-0.03r)$, where $r$ is in units of au. Their model predicts that the planet sculpting this disc will have a semi-major axis of 133 au. By scaling up our simulated disc from its peak of 5.675 au to Fomalhaut's peak of 144 au, we also predict the location of the underlying planet to be 133 au.}
\end{figure}

The existence of a circumstellar disc around the 440 Myr old A3V star Fomalhaut has been known for a long time due to the infrared excess in its spectrum. Fomalhaut is one of the best-studied debris discs, due to its distance from Earth being only 7.7 pc and the fact that it is one of the closest systems that is not edge on. The Fomalhaut debris disc was first directly imaged by \citet[][]{K05} using the Hubble Space Telescope, revealing a sharp inner edge and the central star being offset from the disc geometric centre.
 They derived a flux profile for Fomalhaut by fitting a scattered light model of an inclined, optically thin, belt of dust to observational data, as shown in Fig. \ref{kalas}. The best-fitting value for the power law profile describing the inner edge of the belt is proportional to $r^{10.9}$ whereas that of the outer belt scales with $r^{-4.6}$. Similarly, when an exponential profile is used, the inner edge is proportional to $\exp(0.08r)$ while the outer edge scales with $\exp(-0.03r)$, where $r$ is in au. They define the inner edge as 120--140 au and the outer edge as 140--158 au. 
 
 Since \citet[][]{K05} did not explicitly measure a ring width for Fomalhaut, we extrapolate one from their distribution by defining Fomalhaut's normalized ring width to be equal to its full width at half maximum (FWHM) divided by the peak. Under this definition, Fomalhaut's flux profile has a ring width of $\Delta R / R = 0.191$,  which fits 
 the definition from \citet[][]{HDM18} of a `narrow' ring as one having a normalized ring width $\Delta R / R < 0.5$.

In order to place the Kalas observations within our paradigm, we fit our function to observations of the radial distribution of the  data for Fomalhaut obtained by \citep[][]{K05}. In Fig. \ref{kalas}, the red data points are the raw data obtained by Hubble Space Telescope observations, the blue curve is the fit performed by \citet[][]{K05}, and the black curve is our function's fit to the same data. If the Fomalhaut debris disc were formed by a hypothetical planet's irregular satellite cloud, we can estimate that planet's semi-major axis by scaling up our simulated system. Specifically, we scale up the simulated system from Fig. \ref{rad_prof_B_0_15} so that the peak of the simulated disc's radial profile (5.675 au) matches the peak of the Fomalhaut debris disc's radial profile (144 au). Our model predicts that the planet feeding this disc would have a semi-major axis of 133 au.  We fit the piecewise function to the radial distribution to characterize both the inner edge and outer edge. However, our inner edge is best described by a single-sided Gaussian as opposed to either a power law or an exponential tail. We find that the inner edge has a characteristic length of 2.23 au and that the outer edge has an exponential decay scale of 20.6 au. The inner edge behavior is very similar to what \citet[][]{K05} found in their fit for Fomalhaut, but our function has a more gently-sloped outer edge than their fit. If we measure the ring width of the Fomalhaut disc, we get $\Delta R / R = 0.215$, not significantly different from that of \citet[][]{K05}. Thus, the observed profile of the Fomalhaut debris disc is well fit by that expected for an ISDD.

As a reminder, the thickness of the forbidden zone from the restricted three-body problem governs not only the thickness of any ring gaps that form, but also defines the size of the offset between the peak of the radial distribution and the location of the planet \citep[e.g.,][]{C09,W80}. Such a prediction can be made because the planet is located approximately halfway between the inner and outer edge of the forbidden zone, as shown in Fig. \ref{forbidden_zone_thickness_example}, for example. In the case of a finite $\beta$, we expect the peak of the radial distribution to occur just outside of the outer edge of the forbidden zone. Such a phenomenon should occur because the outer edge of the forbidden zone is also a zero-velocity curve upon which a dust grain decelerates and comes to rest in the co-rotating frame, thus statistically spending more time near the forbidden zone. Therefore, we expect the planet to be a distance of one-half of the forbidden zone thickness interior to the location of the peak.

\begin{figure}
\centering
\includegraphics[width=0.5\textwidth]{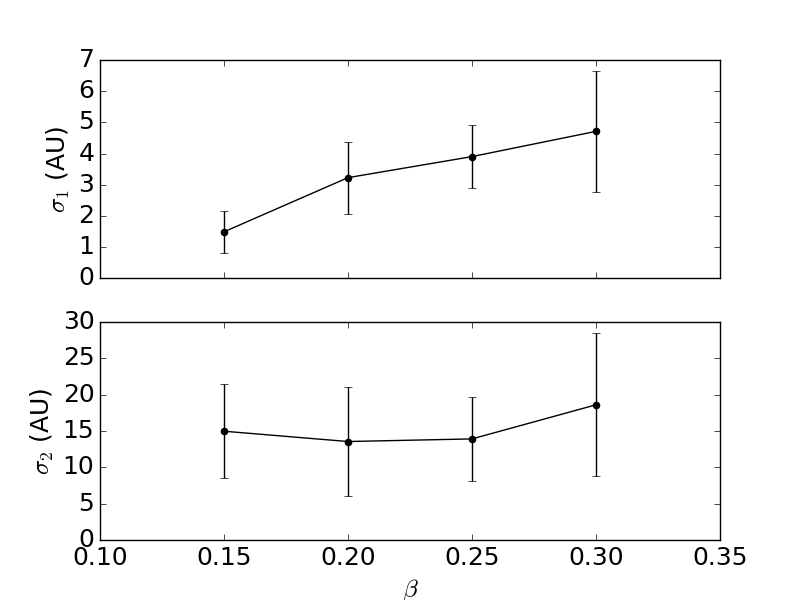}
\caption{\label{sigma_vs_B} Inner edge characteristic length and outer edge characteristic length for various values of $\beta$ for a system with a 1 $\mathrm{M_\odot}$ star and a $10^{-3}\ \mathrm{M_\odot}$ planet. These simulations have the same initial conditions as the simulations shown in Fig. \ref{FWHM}. The inner edge, $\sigma_1$, generally has greater lengths with increasing $\beta$. Interestingly, the outer edge, $\sigma_2$, generally has relatively constant length as a function of $\beta$. However, the sum of $\sigma_1$ and $\sigma_2$ shows that ring width increases as a function of $\beta$. The data begin to become unreliable and noisy at $\beta = 0.3$ due to a small surviving sample size. }
\end{figure}

In order to physically interpret our model fit to the Fomalhaut observations, we first estimate which value of $\beta$, for a single uniform-$\beta$ debris disc, best corresponds to the parameters derived from our fit. We start by plotting two key parameters, $\sigma_1$ and $\sigma_2$, as functions of $\beta$, as shown in Fig. \ref{sigma_vs_B}. These two parameters were chosen because they directly determine the measured width of the ring. As one can see, both the inner edge characteristic length and outer edge characteristic length become larger with increasing $\beta$. The relatively flat distributions show that our model is quite robust and can replicate Fomalhaut observations for a wide variety of radiation pressure strengths, specifically $\beta \leq 0.3$. This is possible because while there is a very weak dependence of $\sigma_2$ on $\beta$, $\sigma_1$ gets much larger at larger $\beta$.


\subsubsection{Fomalhaut b}


 In addition to the debris disc ring, \citet[][]{K08} also detected a point source that was proposed as Fomalhaut~b, a Saturn-mass planet responsible for sculpting the inner edge of the debris disc. In this scenario \citep[][]{C09}, the planet would have a semi-major axis $\sim 115$ au and the inner edge of the dust ring would trace the edge of the chaotic region surrounding the planetary orbit.
 This claim was controversial because the colours of Fomalhaut~b showed little evidence for thermal emission from a giant planet, and were far more consistent with pure scattered light from the star. The reality of the source detection itself has been independently confirmed \citep[][]{C12,G13} but further observations by \citet[][]{K13} reveal several orbital features that make the sculpting planet hypothesis unlikely. The orbit of Fomalhaut~b appears to be highly eccentric ($e \sim 0.8$), especially compared to the eccentricity of the disc ($e = 0.12 \pm 0.01$) \citep[][]{MacGregor17}, so that it would pass through the debris disc if it were not inclined at $\sim 17^\circ$ to the disc. A planet on such an orbit would be unlikely to gravitationally sculpt the observed structure, as the high eccentricity and nonzero inclination does not correspond to the correct orbital geometry to maintain the original model.
 Nor is an object on this orbit likely to be the source of an irregular satellite debris disc, at least according to our model, due to the disparities in both eccentricity and semi-major axis. \citet[][]{K13} estimates the semi-major axis of Fomalhaut b to be 177 au, much larger than that of the planet we propose, which would be  located at 133 au (though their margin of error is quite large at $\pm$68 au). 

 In order to explain the colours of the original Fomalhaut~b hypothetical planet, \citet[][]{KW10} developed a model starting from a similar hypothesis as ours. They constructed a collisional cascade of irregular satellites within a fraction of the Hill sphere of a giant planet, taking into account the strength versus self-gravity of the satellite. They took into account both radiation pressure and Poynting-Robertson (PR) drag for the  resulting dust grains. \citet[][]{KW10} focussed on the appearance of dust confined within the Hill sphere, as a source population for the scattered light observed from Fomalhaut~b. In our model, we focus on the dust that has escaped into heliocentric orbit, as the origin of the debris disc itself -- not the point source.

\subsection{HR 4796A}


HR 4796A is an 8 Myr old A0V star that hosts a well-studied debris disc at a distance of $72.8$ pc from Earth. The disc has an exceptionally high infrared excess of $f = L_{\mathrm{IR}} / L_* = 4.2 \times 10^{-3}$ \citep[][]{J91}. HR 4796A has been imaged in multiple wavelengths including the sub-mm, the mm, mid-infrared, near-infrared, and visible. Combining these different wavelength regimes permits extensive modelling of the spectral energy distribution (SED) of the system. A complete understanding of the SED leads to understanding of the underlying dust composition of the disc. Previous studies have resolved a circular disc structure with a radius of $\sim 77$ au, with a sharply peaked radial profile, and a $\sim 1$ au offset from the location of the star. We can learn more about the dynamics of the system from detailed modeling of the exact geometry.

\subsubsection{HR 4796A ring width}

In 2009, the Hubble Space Telescope resolved the debris disc around HR 4796A and found that it has a ring width of 18.5 au and a radius of 76 au \citep[][]{S09}. Thus, its normalized ring thickness is $\Delta R / R = 0.25$, comparable to that of Fomalhaut and our simulated disc. All three are well within the definition of \citet[][]{HDM18} for a narrow ring.

\begin{figure}
\centering
\includegraphics[width=0.5\textwidth]{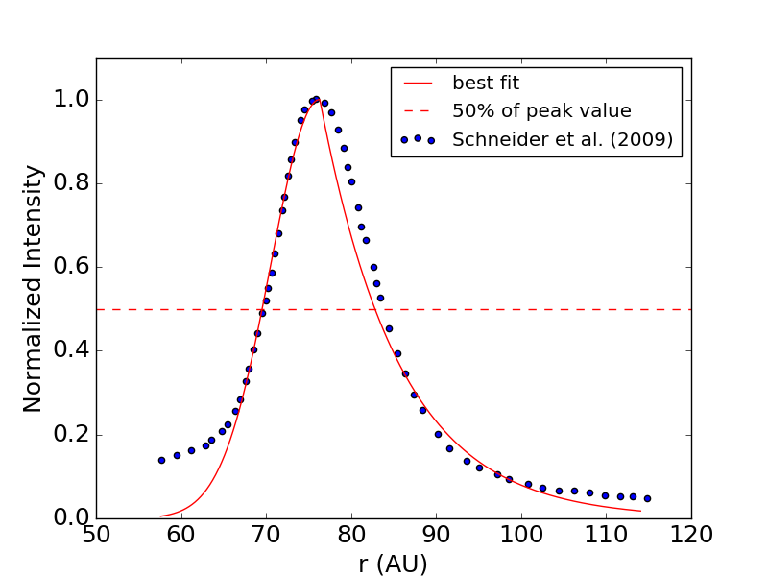}
\caption{\label{arriaga_comparison}  The best fit of our three-regime piecewise function fitted to an HR 4796A flux profile from \citet[][]{S09}. A dashed red line is plotted to indicate 50 per cent of the peak value to help visualize the FWHM. We find that the normalized ring width of the system is $\Delta R / R = 18.3$ per cent. Our model predicts that the planet sculpting this disc will have a semi-major axis of 70.8 au.  }
\end{figure}

We compare our model to observations of HR 4796A made by \citet[][]{S09} using the Hubble Space Telescope Imaging Spectrograph. Specifically, we fit our three-regime piecewise function to the intensity profile for a direct one-to-one comparison and extract a normalized ring width, as shown in Fig. \ref{arriaga_comparison}. If the HR 4796A debris disc were formed by a hypothetical planet's irregular satellite cloud, we can estimate that planet's semi-major axis by scaling up our simulated system. Specifically, we scale up the simulated system from Fig. \ref{rad_prof_B_0_15} so that the peak of the simulated disc's radial profile (5.62 au) matches the peak of the HR 4796A debris disc's radial profile (76.5 au). Our model predicts that the planet feeding this disc will have a semi-major axis of 70.8 au.  We find that our function does an overall good job of fitting the inner edge of the disc, but falls to zero more quickly than the HST data. Mathematically, our function drops to zero quickly since the inner edge is defined by a single-sided Gaussian. The discrepancy may be due to background noise in the HST data. As for the outer edge, our function initially drops off a little bit more quickly than the HST data. As a result, the normalized ring width is slightly lower than that derived from the HST observations, but there is not a significant difference. We once again used the full width at half maximum (FWHM) of the radial profile method for determining ring width and find that the ring with for HR 4796A is $\Delta R/ R = 18.3$ per cent. We note that HR 4796A, Fomalhaut, and our simulated disc all fall within the definition of a `thin ring' as defined by \citet[][]{HDM18}.



\subsubsection{HR 4796A blowout size}

We compare the particle sizes predicted by our model to those predicted by other models for the HR 4796A system. We can calculate the dust grain size corresponding to $\beta = 0.1$ for the stellar parameters of HR 4796A, namely the luminosity of 23 $\mathrm{L_\odot}$ and mass of 2.18 $\mathrm{M_\odot}$, using Equation \ref{beta}. This specific value of $\beta$ was chosen because it fulfills the criterion laid out in Appendix \ref{Threshold} to ensure overflow through the L$_2$ Lagrange point. Rearranging Equation \ref{beta} to solve for the blowout size $D_{\mathrm{bl}}$, we obtain

\begin{equation}
    D_{\mathrm{bl}} \approx (21.4\ \mathrm{\mu m}) \left( \dfrac{\beta}{0.1} \right)^{-1} \left( \dfrac{L_*}{23\ \mathrm{L_\odot}} \right) \left( \dfrac{M_*}{2.18\ \mathrm{M_\odot}} \right)^{-1}.
\end{equation}

The result gives us a dust grain diameter of $D \approx 21.4 \ \mathrm{\mu m}$. \citet[][]{C20} derived a similar grain size of 25 $\mu$m by using MCFOST on SPHERE SPF data. \citet[][]{M17} found the grain size range 17.8--30 $\mu$m fit the data depending on the exact scattering model used.

\begin{table}
\centering
\begin{tabular}{|c|c|c|}
\hline
Division name & Wavelength & $\beta$ \\ \hline \hline 
Mid-wavelength infrared & 3 - 8 $\mu$m & 0.21 - 0.55 \\ \hline 
Long-wavelength infrared & 8 - 15 $\mu$m & 0.11 - 0.21 \\ \hline 
Far infrared & 15 $\mu$m - 1 mm & 0.002 - 0.11 \\ \hline
\end{tabular}
\caption{\label{Wavelengths} Summary of correspondence between observing wavelength and relative strength of radiation pressure $\beta$.}
\end{table}

A general rule of thumb states that dust grains are best observed in electromagnetic radiation at wavelengths that are approximately equal to their size \citep[e.g.,][]{HDM18}. This phenomenon  can be explained as a balance between two opposing processes. We first consider the fact that the smallest grains dominate the grain size distribution and thus contribute the most to the total cross section. However, there is a competing effect where grains can only efficiently emit at wavelengths that are smaller than their actual size. A sharp cutoff in this emission efficiency occurs at larger wavelengths. All in all, the total light emitted will receive contributions from the smallest grains that are able to efficiently emit from that wavelength. The observing wavelengths versus $\beta$ are highlighted in Table \ref{Wavelengths}. Although debris discs were traditionally detected using IR excesses, we are interested in comparing our simulated discs to scattered light images since IR excesses do not give us a geometric picture. Imaging capabilities can vary strongly amongst the different bands in Table \ref{Wavelengths}.


\subsection{Dependence on planet mass}

Generally speaking, we expect the morphologies of ISDDs to depend on the ratio of planet mass to stellar mass $M_\mathrm{p} / M_*$. This dependence arises because the Jacobi contours of the restricted three-body problem depend only on the ratio of the masses of the secondary body to the primary body $M_2 / M_1$, both with and without radiation pressure. For example, a Saturn-like planet orbiting around an M dwarf could have the same normalized Jacobi contours as a Jupiter-like planet orbiting around a G-type star. 

Since the thickness of the Jacobi forbidden zone is roughly the same size as the diameter of the Hill sphere, we expect any ensuing ring gaps to be a similar size to the Hill diameter as well. This effect would likely only be noticeable in low radiation pressure scenarios, since those are the only cases where a significant number of dust grains escape interior to the planet's orbit and would therefore produce an observable ring gap in the radial distribution. Appendix \ref{forbidden_zone_thickness_appendix} goes into greater detail the theoretical foundation of exactly how the mass ratio correlates with the critical $\beta$.

\subsection{General Predictions for Other Systems}

\subsubsection{Wavelength Dependence}

Different observing wavelengths will be able to probe different structures of potentially the same debris disc. However, in any given system, grain size is just a guideline for observing wavelength. Crudely, wavelength corresponds to grain size, so we do have broad predictions about how things should appear. For example, since observing wavelength is expected to be directly proportional to grain size and therefore inversely proportional to $\beta$, we predict that the long-wavelength infrared ($\lambda = 8-15\ \mathrm{\mu m}$) will find singular thin rings. Due to the stronger influence of radiation pressure, mid-wavelength infrared ($\lambda = 3-8\ \mathrm{\mu m}$) will detect comparatively broader rings than found in the long-wavelength infrared, but still one singular ring. However, in the far infrared ($\lambda = 15\ \mathrm{\mu m} - 1\ mm$), which can be affected by values of $\beta$ as low as 0.002, we expect there to be a gap in the ring since Roche lobe overflow through the L$_1$ Lagrange point is almost equally favorable to occur energetically. In this instance, dust escapes both inward and outward through L$_1$ and L$_2$, but the Jacobi forbidden zone prevents the two populations from mixing, giving rise to the gap in the ring. 

We calculate in Appendix \ref{circumstellar} that an initial irregular satellite population mass on the order of 1 $\textrm{M}_\oplus$ is necessary to ensure that the Hill sphere collisional time-scale ($\sim$135  Myr) is less than the age of the age of the system, 440 Myr in the case of Fomalhaut. While 1 $\textrm{M}_\oplus$ of irregular satellites does seem quite large compared to the estimated $10^{-3}$ $\textrm{M}_\textrm{L}$ worth of irregular satellites estimated to have been found around each giant planet in our Solar System \citep[][]{Bott10}, a Jupiter-mass planet found at a semi-major axis consistent with the scale of the Fomalhaut system ($\sim$140 au) orbiting around a 2 $\textrm{M}_\odot$ star would have a Hill sphere $\sim 4 \times 10^4$ times more voluminous than that of Jupiter and a commensurate cross section.

\subsubsection{Evolutionary Implications}

At first glance, irregular satellite collisions would appear to not explain the debris discs found around such aged systems as Fomalhaut (440 Myr), due to how quickly they grind down to dust and dissipate ($\sim$tens to hundreds of thousands of years). Thus, irregular satellite debris discs would only be bright enough to be detectable in their infancy. However, irregular satellites are not formed in the same way as regular satellites, and thus do not have the same age as their host planet. In our own Solar System, they are thought to be the result of dynamical capture during late-stage rearrangement of the giant planet orbits \citep{NVM07}, hundreds of Myr after the formation of the Solar System. This delay may help us explain the age of some older debris disc systems. Our model is not intended to explain every debris disc, but is focussed on the curious geometry of the thin ring systems. Within the proposed context, the observed thin rings indicate systems that have recently emerged from a period of dynamical excitation which resulted in the capture of irregular satellites around giant planets.
 
 The size of the dust in such discs may also be a function of time, because the $\beta$ of the escaping material is set by the balance between residence and collision times. The latter will increase as the mass reservoir in the source population grinds down, moving the characteristic $\beta$ of the escaping particles to lower values, and therefore increasing the size of the particles in the disc.

\section{Conclusions}


%
In this paper, we explored the effects of including radiation pressure into the classical restricted three-body problem. We found that the traditional Roche lobe overflow can be replaced by Lagrange point L$_2$ overflow for a sufficiently high $\beta$ for a given planet-to-star mass ratio $\mu_2 / \mu_1$. Sample orbital integrations reveal that individual dust grains typically trace out `flower petal' orbits, coming to rest on the zero-velocity curves for some time. 

We assumed that the origin of dust grains in our model were from collisions between the giant planet's irregular satellites. We motivated our initial conditions based off of observations of the Solar System giant planets' irregular satellites today as well as what previous studies' determined from their dynamical history. We describe the size distribution of bodies ensuing from irregular satellite collisions as a collisional cascade power law distribution. We calculate the catastrophic collisional time-scale and compare it to an empirically determined residence time-scale to determine the critical $\beta$ at which ground down dust grains can escape the Hill sphere.

Our N-body simulations show that dust grains with a $\beta$ above $\beta_{\mathrm{crit}}$ quickly escape from the Hill sphere and transition from a circumplanetary orbit to a circumstellar orbit. After a short time, a large population of dust grains achieve an azimuthally symmetric disc appearance. We evaluated this azimuthal symmetry by comparing the fluctuations in the azimuthal profile to the average column density and found that they were low. We also calculated the average radius along a given azimuthal angle $\theta$ and found that the mean and median radius is consistent along all azimuthal angles. 

We fit a piecewise function with an Gaussian inner edge and exponential outer edge to the radial profile. These functions naturally allowed us to quantify the  ring width for various values of $\beta$. We normalized the ring width over the ring radius as is standard in the literature ($\Delta R / R$), and find that normalized ring width broadens as a function of $\beta$. We explain this finding as stronger radiation pressure being able to excite dust grains to more eccentric orbits and therefore broadening the overall distribution. Since the vertical profile of the disc resembles a typical Gaussian, we conclude that the overall shape of the disc is a torus.

We compared our results to observations for the specific systems of Fomalhaut and HR 4796A, but also make general predictions for all systems. For the assumption of uniform density spherical dust grains, there is an inverse relationship between observing wavelength and $\beta$. We find that the topology of the debris disc is dictated by the original Jacobi forbidden zone contours, so the fundamental parameter is the planet-to-star mass ratio $M_2 / M_1$. 

We test the validity of our radial profile fitting function by applying it to the raw Hubble Space Telescope data of Fomalhaut from \citet[][]{K05}. We obtain very similar results to their fit in terms of inner and outer edge slopes. By defining a ring width for Fomalhaut as its full width at half maximum, we measure its normalized ring width to be $0.191$, comparable to our model's $\Delta R / R = 0.13$, both of which are within the `thin ring' definition defined in \citet[][]{HDM18} of $\Delta R /R = 0.5$. We note that there is an ongoing debate about whether Fomalhaut b is a planet or a transient dust cloud and clarify that due to its inclined orbital plane with respect to the disc plane, we do not assume Fomalhaut b is the source of the debris disc in our model, but rather some other underlying hidden planet.

For the assumption of a Sun-like star, we make general predictions about distinctions between observing wavelengths in the mid-wavelength infrared, long-wavelength infrared, and far infrared. We address the fact that while Solar System irregular satellite swarms tend to grind down very quickly on time-scales of tens to hundreds of thousands of years, they can still explain very old, large systems such as Fomalhaut once properly scaled up using Kepler's Third Law since irregular satellites were not expected to be captured until the Late Heavy Bombardment period in our Solar System.

\section*{Acknowledgements}

This research has made use of NASA's Astrophysics Data System. This research was supported by NASA Grant 443820-HN-21577.

\section*{Data Availability Statement}

The data underlying this article will be shared on reasonable request to the corresponding author.

        \bibliographystyle{mnras}
        \bibliography{example}

\newpage

\appendix
\section{Poynting-Robertson Drag, Ejection, and Collisional Time-scales of a Circumstellar Dust Grain}\label{circumstellar}

We ultimately wish to calculate the Poynting-Robertson (PR) drag, ejection, and collisional time-scales of a circumstellar dust grain, and compare these three time-scales to determine which processes need to be taken into account.

\subsection{Poynting-Robertson Drag Time-scale}

\citet[][]{BLS79} analytically calculated the characteristic orbital decay time $t_{PR}$ for a $e = 0$ orbit under the influence of PR drag to be

\begin{equation}
    t_{\textrm{PR}} = (78.6\ \textrm{Myr}) \left( \dfrac{M_*}{M_\odot} \right)^{-1} \left( \dfrac{\beta}{0.1} \right)^{-1} \left( \dfrac{a}{140\ \textrm{au}} \right)^{2} .
\end{equation}

\subsection{Ejection Time-scale}

We define `ejection time-scale' to be the amount of time it takes for the maximum number of dust grains that escape from the Hill sphere to drop to $1/e$ of its original value. With that being said, $t_\text{eject}$ is fundamentally an empirical quantity that is measured from our simulations, as shown in Fig. \ref{PR_drag}. The ejection time-scale is typically on the order of $10^{5}$ years or lower for dust grains originating around a Jupiter-mass planet on a 5.2 au orbit and on the order of $10^{7}$ years or lower for dust grains originating around a Jupiter-mass planet on a 140 au orbit. 

\subsection{Collisional Time-scale of a Circumstellar Dust Grain}

Next, we calculate the collisional time-scale of dust grains in the circumstellar disc, whose volume will be approximated as a torus. However, doing so first requires us to know the number density of dust grains in the torus. We will assume that the total mass of dust in the torus at any given time is

\begin{equation}\label{M_torus}
    M_{\textrm{torus}} = \dot{M} t_{\textrm{res}},
\end{equation}

\noindent where $\dot{M}$ is the rate of mass released from the Hill sphere and $t_{\textrm{res}}$ is the residence or collisional time-scale of dust grains in the circumstellar disc. We will approximate $\dot{M}$ as

\begin{equation}\label{M_dot}
    \dot{M} \sim \dfrac{M_{\textrm{tot}}}{t_{\textrm{coll}}},
\end{equation}

\noindent where $M_{\textrm{tot}}$ is the initial total mass of irregular satellites and $t_{\textrm{coll}}$ is the collisional time-scale of an irregular satellite in the Hill sphere of the planet. We will begin by calculating the latter.

\subsubsection{Collisional Time-scale of a 10-km Irregular Satellite in the Hill Sphere}

We can calculate the collisional time-scale of an irregular satellite in the Hill sphere of the planet from the reciprocal of the rate of collisions, $R_{\textrm{coll}}$:

\begin{equation}
    t_{\textrm{coll}} = \dfrac{1}{R_{\textrm{coll}}} = \dfrac{1}{n\sigma v_{\textrm{rel}}},
\end{equation}

\noindent where $n$ is the number density of irregular satellites, $\sigma$ is the cross-sectional area of an irregular satellite, and $v_{\textrm{rel}}$ is the relative velocity between irregular satellites.

We begin by calculating the number density as

\begin{equation}
    n = \dfrac{N}{V} = \dfrac{N}{\dfrac{4\pi}{3} R^3} = \dfrac{3}{4\pi} \dfrac{N}{R^3},
\end{equation}

\noindent where $N$ is the number of irregular satellites, $V$ is the volume they occupy (assumed to be a sphere), and $R$ is the radius of the spherical volume.

If we assume that the total mass of irregular satellites is distributed into equal-mass parent bodies, then we can calculate the number of irregular satellites as

\begin{equation}
    N = \dfrac{M_{\textrm{tot}}}{m_{\textrm{is}}} = \dfrac{M_{\textrm{tot}}}{\dfrac{4\pi}{3}\left( \dfrac{D}{2} \right)^3 \rho_{\textrm{is}}} = \dfrac{6}{\pi} \dfrac{M_\textrm{tot}}{D^3 \rho_\textrm{is}},
\end{equation}

\noindent where $m_{\textrm{is}}$ is the mass of an individual irregular satellite, $D$ is the diameter of an individual irregular satellite, and $\rho_{\textrm{is}}$ is the bulk density of an individual irregular satellites. By plugging in reasonable values, corresponding to a Jupiter-mass planet at 1 au around the solar mass star, $N$ can be expressed as

\begin{equation}
    N = 1,732\ \left( \dfrac{M_{\textrm{tot}}}{10^{-3} M_{\textrm{L}}} \right) \left( \dfrac{D}{30\ \textrm{km}} \right)^{-3} \left( \dfrac{\rho_{\textrm{is}}}{3\ \textrm{g/cm}^3} \right)^{-1}.
\end{equation}

\noindent where $M_{\textrm{L}}$ is a lunar mass. Next, we shall assume that the irregular satellites are found within 40 per cent of the Hill radius, an assumption that is motivated by questions of orbital stability \citep{Bott10}:

\begin{equation}
    R = \dfrac{2}{5} R_H = \dfrac{2}{5} a \left( \dfrac{M_\textrm{p}}{3M_*} \right)^{1/3}.
\end{equation}

Plugging in reasonable values results in the following expression:

\begin{equation}
    R = 0.4 R_\textrm{H} = (4.085 \times 10^{11}\ \textrm{cm}) \left( \dfrac{a}{1\ \textrm{au}} \right) \left( \dfrac{M_\textrm{p}}{M_\textrm{J}} \right)^{1/3} \left( \dfrac{M_*}{M_\odot} \right)^{-1/3},
\end{equation}

\noindent where $a$ is the planet's semi-major axis, $M_\textrm{p}$ is the planet's mass, $M_\textrm{J}$ is a Jupiter mass, $M_*$ is the stellar mass, and $M_\odot$ is a solar mass. Putting everything together, the number density is found to be

\begin{equation}
    n = \dfrac{3}{4\pi} \left[ \dfrac{\dfrac{6}{\pi} \dfrac{M_\textrm{tot}}{D^3 \rho_\textrm{is}}}{\dfrac{2^3}{5^3} a^3 \left( \dfrac{M_\textrm{p}}{3M_*} \right)} \right] = \dfrac{3^3 \cdot 5^3}{2^4 \cdot \pi^2} \left( \dfrac{M_\textrm{tot} M_*}{a^3 D^3 \rho_\textrm{is} M_\textrm{p}} \right).
\end{equation}

Plugging in reasonable values results in the following expression:

\begin{equation}
    \begin{aligned}
        n = (6.07 \times 10^{-33}\ \textrm{cm}^{-3}) & \left( \dfrac{M_{\textrm{tot}}}{10^{-3} M_{\textrm{L}}} \right) \left( \dfrac{D}{30\ \textrm{km}} \right)^{-3} \left( \dfrac{\rho_{\textrm{is}}}{3\ \textrm{g/cm}^3} \right)^{-1} \\
        & \times \left( \dfrac{a}{1\ \textrm{au}} \right)^{-3} \left( \dfrac{M_\textrm{p}}{M_\textrm{J}} \right)^{-1} \left( \dfrac{M_*}{M_\odot} \right).
    \end{aligned}
\end{equation}

Next, we calculate the cross-sectional area by assuming that the irregular satellites are perfect spheres and thus have a circular cross section:

\begin{equation}
    \sigma = \pi \left( \dfrac{D}{2} \right)^2.
\end{equation}

Then we assume that the typical relative velocity between irregular satellites is on the order of the local Keplerian velocity $v_\textrm{kep}$ around the planet:

\begin{equation}
    v_{\textrm{rel}} \sim v_\textrm{kep} = \left( \dfrac{GM_\textrm{p}}{0.4 R_\textrm{H}} \right)^{1/2} = \left( \dfrac{5G}{2a} \right)^{1/2} (3M_*)^{1/6} M_\textrm{p}^{1/3},
\end{equation}

\noindent where $G$ is the Newtonian gravitational constant. We can once again plug in reasonable values to obtain a numerical value for the relative velocity:

\begin{equation}
    v_{\textrm{rel}} \sim (5.569\ \textrm{km/s}) \left( \dfrac{M_\textrm{p}}{M_\textrm{J}} \right)^{1/3} \left( \dfrac{M_*}{M_\odot} \right)^{1/6} \left( \dfrac{a}{1\ \textrm{au}} \right)^{-1/2}.
\end{equation}

Putting it all together, the collisional rate comes out to be

\begin{equation}
    \begin{aligned}
        R_{\textrm{coll}} \sim (2.391 \times 10^{-14}\ \textrm{s}^{-1}) & \left( \dfrac{M_{\textrm{tot}}}{10^{-3} M_{\textrm{L}}} \right) \left( \dfrac{D}{30\ \textrm{km}} \right)^{-1} \left( \dfrac{\rho_{\textrm{is}}}{3\ \textrm{g/cm}^3} \right)^{-1} \\
        & \times \left( \dfrac{a}{1\ \textrm{au}} \right)^{-7/2} \left( \dfrac{M_\textrm{p}}{M_\textrm{J}} \right)^{-2/3} \left( \dfrac{M_*}{M_\odot} \right)^{7/6}
    \end{aligned}
\end{equation}

\noindent and the collisional time-scale comes out to be

\begin{equation}
    t_\textrm{coll} \sim \left( \dfrac{2^7 \pi}{3^{19/6} 5^4} \right) \left( \dfrac{a^{7/2} D \rho_\textrm{is} M_\textrm{p}^{2/3}}{G^{1/2} M_\textrm{tot} M_*^{7/6}} \right).
\end{equation}

Plugging in reasonable values results in the following expression:

\begin{equation}\label{t_coll}
    \begin{aligned}
        t_{\textrm{coll}} \sim (1.33 \times 10^6\ \textrm{yr}) & \left( \dfrac{M_{\textrm{tot}}}{10^{-3} M_{\textrm{L}}} \right)^{-1} \left( \dfrac{D}{30\ \textrm{km}} \right) \left( \dfrac{\rho_{\textrm{is}}}{3\ \textrm{g/cm}^3} \right) \\
        & \times \left( \dfrac{a}{1\ \textrm{au}} \right)^{7/2} \left( \dfrac{M_\textrm{p}}{M_\textrm{J}} \right)^{2/3} \left( \dfrac{M_*}{M_\odot} \right)^{-7/6}.
    \end{aligned}
\end{equation}

\subsubsection{Rate of Dust Generation}

Next we calculate the rate of dust generation $\dot{M}$ by plugging in a suitable value for the total initial mass of irregular satellites $M_{\textrm{tot}}$ into Equation \ref{M_dot}:

\begin{equation}
    \dot{M} \sim \left( \dfrac{3^{19/6} 5^4}{2^7 \pi} \right) \left( \dfrac{G^{1/2} M_\textrm{tot}^2 M_*^{7/6}}{a^{7/2} D \rho_\textrm{is} M_\textrm{p}^{2/3}} \right).
\end{equation}

Plugging in reasonable values results in the following equations:

\begin{equation}
    \begin{aligned}
        \dot{M} \sim (1.756 \times 10^9\ \textrm{g/s}) & \left( \dfrac{M_{\textrm{tot}}}{10^{-3} M_{\textrm{L}}} \right)^2 \left( \dfrac{D}{30\ \textrm{km}} \right)^{-1} \left( \dfrac{\rho_{\textrm{is}}}{3\ \textrm{g/cm}^3} \right)^{-1} \\
        & \times \left( \dfrac{a}{1\ \textrm{au}} \right)^{-7/2} \left( \dfrac{M_\textrm{p}}{M_\textrm{J}} \right)^{-2/3} \left( \dfrac{M_*}{M_\odot} \right)^{7/6}.        
    \end{aligned}
\end{equation}

\subsubsection{Collisional Time-scale between Dust Grains in the Circumstellar Torus}

We will be able to calculate the mass of dust in the torus $M_{\textrm{torus}}$ using Equation \ref{M_torus} if we calculate the residence time of dust grains in the torus. If we assume that dust is removed by collisions, then the residence time is

\begin{equation}\label{t_coll_torus}
    t_{\textrm{res}} = t_{\textrm{coll}}^{\textrm{torus}} = \dfrac{1}{R_{\textrm{coll}}^{\textrm{torus}}} = \dfrac{1}{n_{\textrm{torus}}\ \sigma_{\textrm{dust}}\ v_{\textrm{torus}}},
\end{equation}

\noindent where $t_{\textrm{coll}}^{\textrm{torus}}$ is the collisional time-scale of a dust grain in the circumstellar torus, $R_{\textrm{coll}}^{\textrm{torus}}$ is the collisional rate, $n_{\textrm{torus}}$ is the number density of dust grains in the torus, $\sigma_{\textrm{dust}}$ is the cross-sectional area of a single dust grain, and $v_{\textrm{torus}}$ is the relative velocity between dust grains in the torus.

We begin by calculating the relative velocity by once again assuming it is on the order the Keplerian velocity, but this time with respect to the star:

\begin{equation}
    v_{\textrm{torus}} = f \left( \dfrac{GM_*}{a} \right)^{1/2},
\end{equation}

\noindent where $f = v_z / v_c$ is the ratio of vertical velocity to circular velocity, which is in turn on the order of the ratio of the scale height of the disk $H$ to the width of the disk $\Delta R$.

Plugging in Solar System values gives us the average orbital velocity of the Earth around the Sun as a benchmark:

\begin{equation}
    v_{\textrm{torus}} = (4.468\ \textrm{km/s}) \left( \dfrac{M_*}{M_\odot} \right)^{1/2} \left( \dfrac{a}{1\ \textrm{au}} \right)^{-1/2} \left( \dfrac{f}{0.15} \right).
\end{equation}

We then calculate the cross-sectional area of the dust grains, by assuming that the dust grains are perfect spheres with circular cross sections:

\begin{equation}
    \sigma_{\textrm{dust}} = \pi r^2,
\end{equation}

\noindent where $r$ is the radius of an individual dust grain. Plugging in 50 $\mu$m as a representative value results in

\begin{equation}
    \sigma_{\textrm{dust}} = (7.854 \times 10^{-5}\ \textrm{cm}^2) \left( \dfrac{r}{50\ \mu \textrm{m}} \right)^2.
\end{equation}

We determine the number density of dust grains from $n_{\textrm{torus}} = N_{\textrm{dust}}/V_{\textrm{torus}}$, where $N_{\textrm{dust}}$ is the number of dust grains in the torus and $V_{\textrm{torus}}$ is the volume of the torus. We first estimate the volume of the torus as having a major radius equal to the semi-major axis of the planet and minor radius half the size of the ring width $\Delta a$:

\begin{equation}
    V_{\textrm{torus}} \sim 2\pi a \cdot \pi \left (\dfrac{\Delta a}{2} \right)^2.
\end{equation}

Plugging in reasonable values results in the following expression for the volume:

\begin{equation}
    V_{\textrm{torus}} \sim (5.23698 \times 10^{40}\ \textrm{cm}^3) \left( \dfrac{a}{1\ \textrm{au}} \right)^3 \left( \dfrac{\Delta a / a}{0.15} \right)^2.
\end{equation}

We use Equation \ref{M_torus} to calculate the mass of the torus in terms of the residence time $t_{\textrm{res}}$:

\begin{equation}\label{M_torus}
    M_\textrm{torus} \sim \left( \dfrac{3^{19/6} 5^4}{2^7 \pi} \right) \left( \dfrac{G^{1/2} M_\textrm{tot}^2 M_*^{7/6} t_\textrm{res}}{a^{7/2} D \rho_\textrm{is} M_\textrm{p}^{2/3}} \right).
\end{equation}

If we assume that the torus is uniformly composed of equal sized dust grains, then the total number of dust grains is

\begin{equation}\label{N_dust}
    N_{\textrm{dust}} = \dfrac{M_{\textrm{torus}}}{\dfrac{4\pi}{3} r^3 \rho_{\textrm{dust}}} = \left( \dfrac{3^{25/6} 5^4}{2^9 \pi^2} \right) \left( \dfrac{G^{1/2} M_\textrm{tot}^2 M_*^{7/6} t_\textrm{res}}{a^{7/2} D r^3 \rho_\textrm{is} \rho_\textrm{dust} M_\textrm{p}^{2/3}} \right),
\end{equation}

\noindent where $\rho_{\textrm{dust}}$ is the bulk density of dust grains. 

Putting everything together results in the following expression for number density:

\begin{equation}
    \begin{aligned}
        n_{\textrm{dust}} \sim & (1.01 \times 10^{-19}\ \textrm{cm}^{-3}) \left( \dfrac{t_{\textrm{res}}}{1\ \textrm{yr}} \right) \left( \dfrac{M_{\textrm{tot}}}{10^{-3} M_{\textrm{L}}} \right)^2 \left( \dfrac{D}{30\ \textrm{km}} \right)^{-1} \\
        & \times \left( \dfrac{\rho_{\textrm{is}}}{3\ \textrm{g/cm}^3} \right)^{-1} \left( \dfrac{\rho_{\textrm{dust}}}{3\ \textrm{g/cm}^3} \right)^{-1} \left( \dfrac{a}{1\ \textrm{au}} \right)^{-13/2} \left( \dfrac{M_\textrm{p}}{M_\textrm{J}} \right)^{-2/3} \\
        & \times \left( \dfrac{M_*}{M_\odot} \right)^{7/6} \left( \dfrac{r}{50\ \mu \textrm{m}} \right)^{-3} \left( \dfrac{\Delta a / a}{0.15} \right)^{-2} \left( \dfrac{f}{0.15} \right).
    \end{aligned}
\end{equation}

Lastly, we multiply both sides by $t_{\textrm{res}}$ then take the square root to isolate $t_{\textrm{res}}$:

\begin{equation}
    t_\textrm{res} \sim \left( \dfrac{2^5 \pi}{3^{25/12} 5^2} \right) \left( \dfrac{a^{7} D r \rho_\textrm{is} \rho_\textrm{dust} M_\textrm{p}^{2/3}}{G M_\textrm{tot}^2 M_*^{5/3} f} \right)^{1/2} \left( \dfrac{\Delta a}{a} \right)
\end{equation}

Plugging in reasonable values results in the following:

\begin{equation}
    \begin{aligned}
        t_{\textrm{res}} \sim & (36,623\ \textrm{yr}) \left( \dfrac{M_{\textrm{tot}}}{10^{-3} M_{\textrm{L}}} \right)^{-1} \left( \dfrac{D}{30\ \textrm{km}} \right)^{1/2} \left( \dfrac{\rho_{\textrm{is}}}{3\ \textrm{g/cm}^3} \right)^{1/2} \left( \dfrac{a}{1\ \textrm{au}} \right)^{7/2} \\
        & \times \left( \dfrac{\rho_{\textrm{dust}}}{3\ \textrm{g/cm}^3} \right)^{1/2} \left( \dfrac{M_\textrm{p}}{M_\textrm{J}} \right)^{1/3} \left( \dfrac{M_*}{M_\odot} \right)^{-5/6} \left( \dfrac{r}{50\ \mu \textrm{m}} \right)^{1/2} \left( \dfrac{\Delta a / a}{0.15} \right) \left( \dfrac{f}{0.15} \right)^{-1/2}.
    \end{aligned}
\end{equation}

Since $\beta$ is defined to be

\begin{equation}
    \beta = \dfrac{3L_* \langle Q_\textrm{rad} \rangle}{8 \pi GM_* c \rho_{\textrm{dust}} (2r)},
\end{equation}

\noindent where $L_*$ is the stellar luminosity, $\langle Q_\textrm{rad} \rangle$ is the wavelength-averaged radiation pressure coefficient, and $c$ is the speed of light, we can create a one-to-one correspondence between $r$ and $\beta$:

\begin{equation}
    r = (3.207\ \mu \textrm{m}) \left( \dfrac{\beta}{0.03} \right)^{-1} \left( \dfrac{M_*}{M_\odot} \right)^{-1} \left( \dfrac{L_*}{L_\odot} \right) \left( \dfrac{\rho_{\textrm{dust}}}{3\ \textrm{g/cm}^3} \right)^{-1} \left( \dfrac{\langle Q_\textrm{rad} \rangle}{0.5} \right).
\end{equation}

We can express $t_{\textrm{res}}$ in terms of $\beta$ instead of $r$ as

\begin{equation}
    t_\textrm{res} \sim \left( \dfrac{2^3 \pi^{1/2}}{3^{19/12} 5^2} \right) \left( \dfrac{a^{7} D \rho_\textrm{is} M_\textrm{p}^{2/3} L_* \langle Q_\textrm{rad} \rangle}{G^2 M_\textrm{tot}^2 M_*^{8/3} f c \beta} \right)^{1/2} \left( \dfrac{\Delta a}{a} \right).
\end{equation}

Plugging in reasonable values results in the following equation:

\begin{equation}
    \begin{aligned}
        t_{\textrm{res}} = (700\ \textrm{yr}) 
        &  \left( \dfrac{M_{\textrm{tot}}}{10^{-3} M_{\textrm{L}}} \right)^{-1} \left( \dfrac{\beta}{0.03} \right)^{-1/2} \left( \dfrac{\rho_{\textrm{is}}}{3\ \textrm{g/cm}^3} \right)^{1/2} \left( \dfrac{L_*}{L_\odot} \right)^{1/2} \\
        & \times \left( \dfrac{\langle Q_\textrm{rad} \rangle}{0.5} \right)^{1/2} \left( \dfrac{a}{1\ \textrm{au}} \right)^{7/2} \left( \dfrac{M_\textrm{p}}{M_\textrm{J}} \right)^{1/3} \left( \dfrac{M_*}{M_\odot} \right)^{-4/3} \\
        & \times \left( \dfrac{D}{30\ \textrm{km}} \right)^{1/2} \left( \dfrac{\Delta a / a}{0.15} \right) \left( \dfrac{f}{0.15} \right)^{-1/2}.
    \end{aligned}
\end{equation}

When we use Kepler's Third Law to scale up to $a = 140$ au to represent a Fomalhaut-like system, this equation becomes:

\begin{equation}\label{t_res_fom}
    \begin{aligned}
        t_{\textrm{res}} = (22.7\ \textrm{Myr}) & \left( \dfrac{M_{\textrm{tot}}}{M_{\textrm{L}}} \right)^{-1} \left( \dfrac{\beta}{0.1} \right)^{-1/2} \left( \dfrac{\rho_{\textrm{is}}}{3\ \textrm{g/cm}^3} \right)^{1/2} \\
        & \times \left( \dfrac{L_*}{L_\odot} \right)^{1/2} \left( \dfrac{\langle Q_\textrm{rad} \rangle}{0.5} \right)^{1/2} \left( \dfrac{a}{140\ \textrm{au}} \right)^{7/2} \left( \dfrac{M_\textrm{p}}{M_\textrm{J}} \right)^{1/3} \\
        & \times \left( \dfrac{M_*}{M_\odot} \right)^{-4/3} \left( \dfrac{D}{30\ \textrm{km}} \right)^{1/2} \left( \dfrac{\Delta a / a}{0.15} \right) \left( \dfrac{f}{0.15} \right)^{-1/2}.
    \end{aligned}
\end{equation}

Now that $t_\textrm{res}$ has been isolated and solved for in terms other than itself, we can go back and calculate the total mass of the disc by plugging Equation \ref{t_res_fom} into Equation \ref{M_torus}, we obtain the following expression:

\begin{equation}
    \begin{aligned}
        M_\textrm{torus} & \sim (1.75 \times 10^{16}\ \textrm{kg}) \left( \dfrac{M_{\textrm{tot}}}{10^{-3}\ M_{\textrm{L}}} \right)^{-1} \left( \dfrac{\beta}{0.03} \right)^{-1/2} \\
        & \times \left( \dfrac{\rho_{\textrm{is}}}{3\ \textrm{g/cm}^3} \right)^{-1/2} \left( \dfrac{L_*}{L_\odot} \right)^{1/2} \left( \dfrac{\langle Q_\textrm{rad} \rangle}{1} \right)^{1/2} \left( \dfrac{M_\textrm{p}}{M_\textrm{J}} \right)^{-1/3} \\
        & \times \left( \dfrac{M_*}{M_\odot} \right)^{-1/6} \left( \dfrac{D}{30\ \textrm{km}} \right)^{-1/2} \left( \dfrac{\Delta a / a}{0.15} \right) \left( \dfrac{f}{0.15} \right)^{-1/2}.
    \end{aligned}
\end{equation}

Additionally, we can go back and similarly obtain an expression for the total number of dust grains in the disc by plugging Equation \ref{t_res_fom} into Equation \ref{N_dust}:

\begin{equation}
    \begin{aligned}
        N_\textrm{dust} & \sim (1.499 \times 10^{29}) \left( \dfrac{M_{\textrm{tot}}}{10^{-3}\ M_{\textrm{L}}} \right) \left( \dfrac{\beta}{0.03} \right)^{5/2} \left( \dfrac{\rho_{\textrm{dust}}}{3\ \textrm{g/cm}^3} \right)^{2} \\
        & \times \left( \dfrac{\rho_{\textrm{is}}}{3\ \textrm{g/cm}^3} \right)^{-1/2} \left( \dfrac{L_*}{L_\odot} \right)^{-5/2} \left( \dfrac{\langle Q_\textrm{rad} \rangle}{1} \right)^{-5/2} \left( \dfrac{M_\textrm{p}}{M_\textrm{J}} \right)^{-1/3} \\
        & \times \left( \dfrac{M_*}{M_\odot} \right)^{17/6} \left( \dfrac{D}{30\ \textrm{km}} \right)^{-1/2} \left( \dfrac{\Delta a / a}{0.15} \right) \left( \dfrac{f}{0.15} \right)^{-1/2}.
    \end{aligned}
\end{equation}

We see in Eq. \ref{t_res_fom} that the collisional time-scale $t_\textrm{res}$ is large compared to the empirical residence time of dust grains \textit{as shown in Fig. \ref{PR_drag}}, so we assume that the ejection from the system is the dominant form of dust grain loss. 

Both of the two previous expressions represent upper limits since the total disc mass and total number of dust grains are only independent of semi-major axis when collisionally limited.

\subsection{Fractional Luminosity of an ISDD}

{We next calculate the fractional luminosity of an ISDD to quantify the amount of infrared emission from our circumstellar discs and further assess the feasibility of irregular satellite collisions as the progenitors of thin extrasolar debris rings. Fractional luminosity is typically defined as the ratio of dust grain bolometric luminosity to stellar bolometric luminosity, or $f = L_\text{dust} / L_*$. For the purposes of this order-of-magnitude estimate, we will assume that the disc is homogeneously composed of uniform-size dust grains and that such grains are perfect blackbodies, so}

\begin{equation}\label{frac_lum}
    f = \dfrac{L_\text{dust}}{L_*} = \dfrac{N_\text{dust} (4\pi r^2 \sigma_\text{SB} T_\text{eff}^4)}{L_*},
\end{equation}

\noindent {where $\sigma_\text{SB}$ is the Stefan-Boltzmann constant and $T_\text{eff}$ is the effective temperature of a dust grain. In thermal equilibrium, the energy per unit time absorbed by a dust grain must be equal to the energy per unit time emitted by that dust grain, so}

\begin{equation}\label{therm_eq}
    \left( \dfrac{A_\text{dust}}{A_\text{orbit}} \right) L_* = 4\pi r^2 \sigma_\text{SB} T_\text{eff}^4,
\end{equation}

\noindent {where $A_\text{dust} = \pi r^2$ is the cross-sectional area of a dust grain and $A_\text{orbit} = 4\pi a^2$ is the surface area of the sphere with the same radius as the ring, $a$, so the ratio $A_\text{dust} / A_\text{orbit}$ represents the fractional amount of radiation intercepted by a dust grain at a distance $a$ from the star. Combining Equations \ref{frac_lum} and \ref{therm_eq} results in}

\begin{equation}
    f = \dfrac{N_\text{dust}}{16} \left( \dfrac{r}{a} \right)^2.
\end{equation}

{When we plug in the expression for $N_\text{dust}$ in the previous subsection, we obtain the following analytic expression:}

\begin{equation}
    f \sim \left( \dfrac{3^{19/12} \cdot 5^2}{2^6 \cdot \pi^{1/2}} \right) \left( \dfrac{M_\text{tot} G^{1/2} M_*^{5/6} c^{1/2} \beta^{1/2}}{L_*^{1/2} \langle Q_\text{rad} \rangle^{1/2} D^{1/2} \rho_\text{is}^{1/2} M_\text{p}^{1/3} f_\text{V}^{1/2} a^2} \right) \left( \dfrac{\Delta a}{a} \right),
\end{equation}

\noindent {and the following scaling relationship once reasonable values for a Jupiter-mass planet in a Solar System-like configuration are plugged in:}

\begin{equation}
    \begin{aligned}
        f  \sim (4.27 \times 10^{-6}) & \left( \dfrac{M_\text{tot}}{10^{-3}\ M_\text{L}} \right) \left( \dfrac{M_*}{M_\odot} \right)^{5/6} \left( \dfrac{\beta}{0.03} \right)^{1/2} \left( \dfrac{L_*}{L_\odot} \right)^{-1/2} \\
        & \times \left( \dfrac{\langle Q_\textrm{rad} \rangle}{0.5} \right)^{-1/2} \left( \dfrac{D}{30\ \text{km}} \right)^{-1/2} \left( \dfrac{\rho_\textrm{is}}{3\ \textrm{g/cm}^3} \right)^{-1/2} \\
        & \times \left( \dfrac{M_\text{p}}{M_\text{J}} \right)^{-1/3} \left( \dfrac{f_\textrm{v}}{0.15} \right)^{-1/2} \left( \dfrac{\Delta a / a}{0.15} \right) \left( \dfrac{a}{1\ \text{au}} \right)^{-2},
    \end{aligned}
\end{equation}

\noindent {where the ratio of vertical velocity to circular velocity has been redefined as $f_\textrm{v}$ to avoid confusion with fractional luminosity $f$. We obtain the following scaling relation for a Jupiter-mass planet in a Fomalhaut-size configuration ($a = 140$ au, $M_\textrm{tot} = 1.2 \ M_\oplus$):}

\begin{equation}
    \begin{aligned}
        f  \sim (9.02 \times 10^{-5}) & \left( \dfrac{M_\text{tot}}{1.2 \ M_\oplus} \right) \left( \dfrac{M_*}{1.92\ M_\odot} \right)^{5/6} \left( \dfrac{\beta}{0.03} \right)^{1/2} \left( \dfrac{L_*}{16.63\ L_\odot} \right)^{-1/2} \\
        & \times \left( \dfrac{\langle Q_\textrm{rad} \rangle}{0.5} \right)^{-1/2} \left( \dfrac{D}{30\ \text{km}} \right)^{-1/2} \left( \dfrac{\rho_\textrm{is}}{3\ \textrm{g/cm}^3} \right)^{-1/2} \\
        & \times \left( \dfrac{M_\text{p}}{M_\text{J}} \right)^{-1/3} \left( \dfrac{f_\textrm{v}}{0.15} \right)^{-1/2} \left( \dfrac{\Delta a / a}{0.15} \right) \left( \dfrac{a}{140\ \text{au}} \right)^{-2},
    \end{aligned}
\end{equation}

\noindent {where $M_\text{tot} = 1.2 \ M_\oplus$ was found by setting the collisional time-scale $t_\text{coll}$ in Eq. \ref{t_coll} to the age of Fomalhaut, $4.4 \times 10^8$ yr \citep[][]{M12}{}{}}.

{According to \citet[][]{M14}, a typical range of fractional luminosities for known observable discs is $10^{-6}$ to $10^{-3}$, so our simulated debris discs are just on the cusp of being observable with current technology. }

\section{Critical Value for L$_1$/L$_2$ crossover}
\label{Threshold}
In the limit of $\beta=0$, the lowest energy threshold for mass loss from the planetary Hill sphere is through the
L$_1$ point, but this switches to the L$_2$ point if $\beta$ is large enough. We wish to quantify the nature of this
transition. We will follow the notation of \citet[][]{MD99}.

\begin{figure}
\centering
\includegraphics[width=0.5\textwidth]{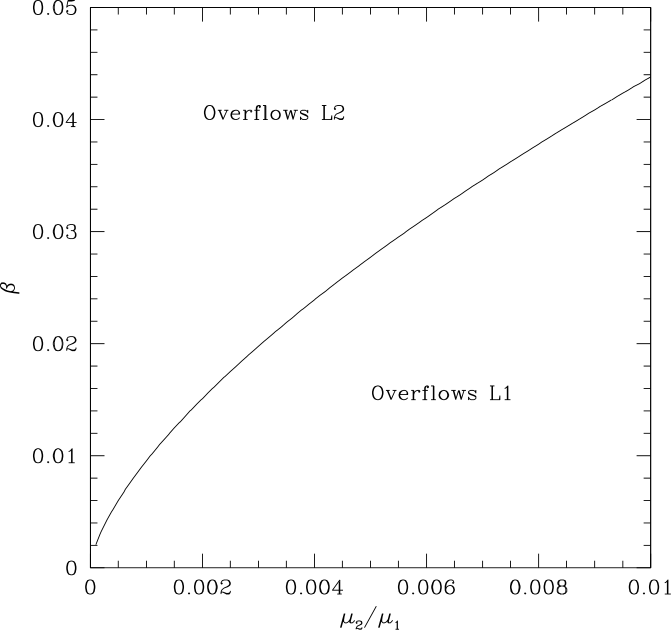}
\caption{\label{ThreshU} The curve illustrates the relationship between the radiation pressure
strength $\beta$ and the planetary mass ratio $\mu_2/\mu_1$, such that the value of the potential
at the L$_1$ and L$_2$ points are the same. For values above this curve, dust particles spiral outwards,
rather than inwards.}
\end{figure}

In the traditional description of the restricted three-body problem, we define a co-ordinate system co-rotating
with two massive bodies, of masses $\mu_1$ and $\mu_2$, such that $\mu_1 + \mu_2=1$. If we further normalise the
angular rotation $n=1$ and designate the x-axis as the one upon which the massive bodies lie, then the dynamical
equations to describe the motion of a test particle in this frame are
\begin{eqnarray}
\ddot{x} - 2\dot{y} - x & = & -\left[ \mu_1 (1-\beta) \frac{(x+\mu_2)}{r_1^3} + \mu_2 \frac{(x-\mu_1}{r_2^3} \right] \\
\ddot{y} + 2\dot{x} - y & = & - \left[ \mu_1 (1-\beta) \frac{y}{r_1^3} + \mu_2 \frac{y}{r_2^3} \right] \\
\ddot{z} &=& - \left[\mu_1 (1-\beta) \frac{z}{r_1^3} + \mu_2 \frac{z}{r_2^3}\right]
\end{eqnarray}
where the massive particles are located at $x=-\mu_2$ and $x=\mu_1$ respectively. As in the case where we ignore
radiation pressure, the dynamics are regulated by a pseudopotential, which takes the form (in the $z=0$ plane)
\begin{equation}
    U = \mu_1 \left[ \frac{1}{2} r_1^2 + \frac{1-\beta}{r_1} \right] + \mu_2 \left[ \frac{1}{2} r_2^2+ \frac{1}{r_2} \right] - \frac{1}{2} \mu_1 \mu_2
\end{equation}
The equilibrium points are to be found at the extrema of this potential. The criterion for the L$_1$ point can be written as
\begin{equation}
    \alpha^3 = r_2^3\frac{\left[1 - r_2 + \frac{1}{3} r_2^2 - \frac{\beta}{3 r_2} \right]}
    {(1+r_2+r_2^2) (1 - r_2)^3}
\end{equation}
where we have used $r_1+r_2=1$, so $r_1=x+\mu_2$ and $r_2=\mu_1-x$. We have also defined $\alpha=[\mu_2/(3\mu_1)]^{1/3}$.
The value of the pseudopotential at this point is
\begin{equation}
 \left( \frac{U_1}{\mu_1} \right) = \frac{1}{2} (1-r_2)^2 + \frac{1-\beta}{1-r_2} + 3 \alpha^3 \left( \frac{1}{2} r_2^2 + \frac{1}{r_2} \right) - \frac{3}{2} \alpha^3    
\end{equation}
One can develop similar expressions for the L$_2$ point, where $r_1=1+r_2$, $r_1=x+\mu_2$ and $r_2=x-\mu_1$. These are
\begin{equation}
    \alpha^3 = r_2^3\frac{\left[1 + r_2 + \frac{1}{3} r_2^2 - \frac{\beta}{3 r_2} \right]}
    {(1+r_2)^2 (1 - r_2)^3}
\end{equation}
\begin{equation}
 \left( \frac{U_2}{\mu_1} \right) = \frac{1}{2} (1+r_2)^2 + \frac{1-\beta}{1+r_2} + 3 \alpha^3 \left( \frac{1}{2} r_2^2 + \frac{1}{r_2} \right) - \frac{3}{2} \alpha^3    
\end{equation}
The topology of the Hill sphere changes when $U_1=U_2$. Fig.~\ref{ThreshU} shows this criterion as a function
of $\beta$ and the mass ratio $\mu_2/\mu_1$. As the planetary mass increases, it requires stronger radiation pressure to generate
an external debris disc.

\section{Forbidden Zone Thickness as a function of radiation pressure and mass ratio}\label{forbidden_zone_thickness_appendix}

In Section \ref{sample_orbital_integrations}, we showed one example of how the forbidden zone contours of the restricted three-body problem changes as a function of radiation pressure by comparing and contrasting the zero-velocity curves for $\beta = 0$ and $\beta = 0.1$. As we saw in that section, there were two primary changes in the zero-velocity curves for that example--the forbidden zone thickness increased with radiation pressure, but the radius of both the inner and outer edges shrunk. In other words, the radius of the outer edge of the forbidden zone shrunk by less than that of the inner edge.

\begin{figure}
\centering
\includegraphics[width=0.5\textwidth]{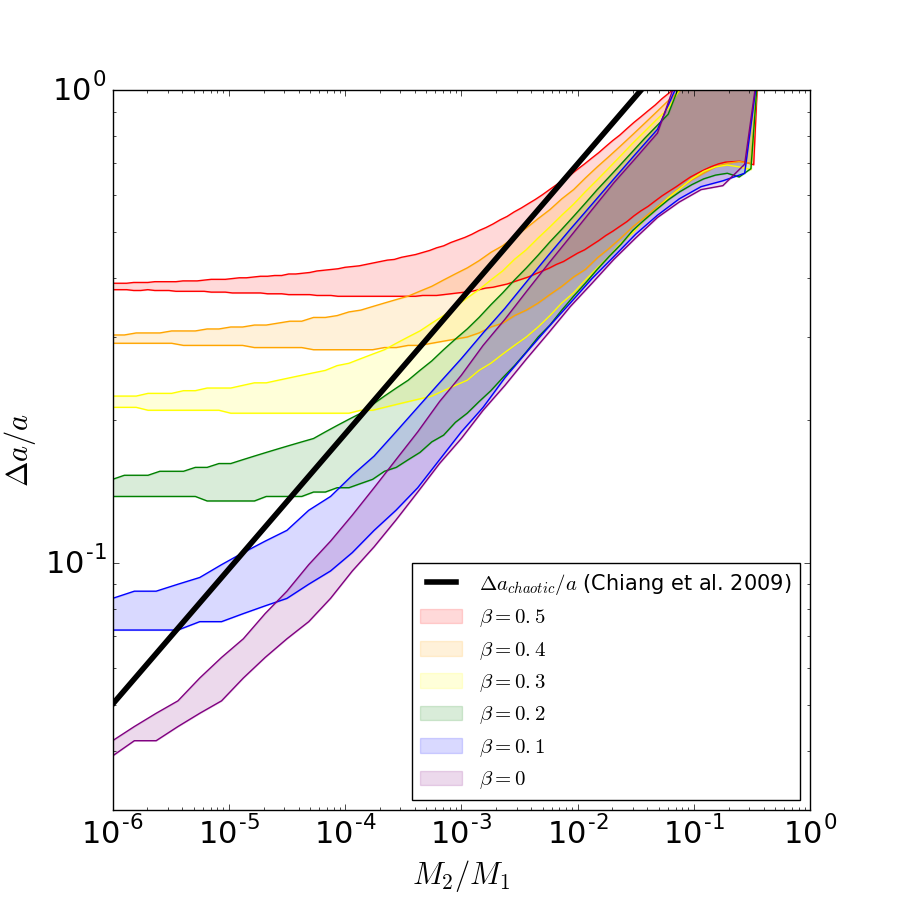}
\caption{\label{forbidden_zone_thickness} Normalized forbidden zone thickness ranges as a function of radiation pressure and mass ratio. The forbidden zone thicknesses have ranges because of the range of Jacobi constants that are allowed arising from the initial conditions ranging from 0.3$R_\mathrm{H}$ to 0.5$R_\mathrm{H}$. }
\end{figure}

In this Appendix, we explore a much more comprehensive range of representative values for radiation pressure strength, $\beta$, and allow the mass ratio {to} vary, as shown in Fig. \ref{forbidden_zone_thickness}. We choose a range of $\beta = 0-0.5$ since $\beta = 0.5$ is the traditional blow-out point in the restricted two-body problem. For mass ratio, we choose a generous range of $M_2/M_1 = 10^{-6}$ to 1 to explore the effects of radiation pressure from as low as an Earth-to-Sun mass ratio up to all possible binary stellar companions. 

The main trend here is that the forbidden zone thickness is much more sensitive to radiation pressure changes for smaller mass ratios. Specifically, small mass ratios are defined as as low as $M_2 / M_1 = 10^{-6}$ like for an Earth-Sun relationship, all the way up to $M_2 / M_1 = 10^{-3}$, like for a Jupiter-Sun relationship. For mass ratios that approach unity for brown dwarf companions and binary stars, radiation pressure does not have much of an effect on forbidden zone thickness. Lastly, we overplot the chaotic zone width results from \citet[][]{C09} to show that we obtained a nearly identical slope to them, just with a slightly different amplitude. Their estimate for the width of the chaotic zone is essentially our upper limit for the width of the forbidden zone, likely due to the fact that we only considered Jacobi constants that originated in the range 0.3$R_\mathrm{H}$ to 0.5$R_\mathrm{H}$.

\bsp	
\label{lastpage}
\end{document}